\documentclass{emulateapj}
\usepackage{graphicx}
\newcommand{\be}{\begin{equation}}
\newcommand{\ee}{\end{equation}}
\newcommand{\bea}{\begin{eqnarray}}
\newcommand{\eea}{\end{eqnarray}}

\begin{document}

\title{Magnetization of cloud cores and envelopes and other observational consequences of reconnection diffusion }
\author{A. Lazarian\altaffilmark{1}, A. Esquivel\altaffilmark{2}, and R. Crutcher\altaffilmark{3}}

\altaffiltext{1}{Astronomy Department, University of Wisconsin, Madison, WI 53706, USA}
\altaffiltext{2}{Instituto de Ciencias Nucleares, Universidad Nacional
Aut\'{o}noma de M\'{e}xico, Apartado Postal 70-543, 04510 M\'{e}xico
D.F., Mexico}
\altaffiltext{3}{Department of Astronomy, University of Illinois at Urbana-Champaign, 1002 W. Green Street,
Urbana, IL 61801, USA}

\begin{abstract}
Recent observational results for magnetic fields in molecular clouds reviewed by Crutcher (2012) seem to be inconsistent with the predictions of the ambipolar diffusion theory of star formation. These include the measured decrease in mass to flux ratio between envelopes and cores, the failure to detect any self-gravitating magnetically subcritical clouds, the determination of the flat PDF of the total magnetic field strengths implying that there are many clouds with very weak magnetic fields, and the observed scaling $B \propto \rho^{2/3}$ that implies gravitational contraction with weak magnetic fields. We consider the problem of magnetic field evolution in turbulent molecular clouds and discuss the process of magnetic field diffusion mediated by magnetic reconnection. For this process that we termed ``reconnection diffusion'' we provide a simple physical model and explain that this process is inevitable in view of the 
present day understanding of MHD turbulence. We address the issue of the expected magnetization of cores and envelopes
in the process of star formation and show that reconnection diffusion provides an efficient removal of magnetic flux that depends only on the properties of MHD turbulence in the core and the envelope. We show that as the amplitude of
turbulence as well as the scale of turbulent motions decrease from the envelope to the core of the cloud,
the diffusion of the magnetic field is faster in the envelope. As a result, the magnetic flux trapped during
the collapse in the envelope is being released faster than the flux trapped in the core, resulting in much weaker fields in envelopes than in cores, as observed. We provide simple semi-analytical model calculations which
support this conclusion and qualitatively agree with the observational results. Magnetic reconnection is also consistent with the lack of subcritical self-gravitating clouds, with the observed flat PDF of field strengths, and with the scaling of field strength with density. In addition, we demonstrate that reconnection diffusion process can account for the empirical Larson (1981) relations and list a few other implications of the reconnection diffusion concept. We argue that magnetic reconnection provides a solution to the magnetic flux problem of star formation that agrees better with observations than the long-standing ambipolar diffusion paradigm. Due to the illustrative nature of our simplified model we do not seek quantitative agreement, but discuss the complementary nature of our approach to the 3D MHD numerical simulations.\end{abstract}


\section{Introduction}
Interstellar media are known to be turbulent and magnetized,  and both turbulence and magnetic field are important for star formation (see Armstrong et al. 1994, Chepurnov \& Lazarian 2009, Crutcher 2012). The existing star formation paradigm has been developed with the concept that near-perfect flux freezing holds,
i.e. that the magnetic field is well coupled with ions and electrons in the media (Alfv\'{e}n 1942). This is assumed in magnetically-mediated star formation theory, which was founded by the pioneering studies by L. Mestel  and L. Spitzer (see Mestel \& Spitzer 1956, Mestel 1966) and  brought to a high level of sophistication by other researchers (see Shu, Adams \& Lizano 1987, Mouschovias 1991, Nakano et al. 2002, Shu et al. 2004, Mouschovias et al. 2006). According to the theory, magnetic fields slow down and even prevent star formation if the media are sufficiently magnetized. The theory {\it assumes} that
the change of the flux to mass ratio happens due to {\it ambipolar diffusion}, i.e. 
to the drift of neutrals which do not feel magnetic fields directly, but only through ion-neutral collisions. Naturally, in the presence of gravity, neutrals get concentrated towards the center of the gravitational potential while magnetic fields resist compression and therefore leave the
forming protostar (e.g. Mestel 1965). The rate of ambipolar diffusion for a cloud in gravitational equilibrium with the magnetic field 
depends only on the degree of ionization of the media. 

The existing theory makes star formation inefficient for magnetically dominated (i.e. subcritical) clouds. The low efficiency of star formation corresponds to observations (e.g. Zuckerman \& Evans 1974), which is usually interpreted as a strong argument in support of the above scenario. This however does not solve all the problems; at the same time, for clouds dominated by gravity, i.e. supercritical clouds, this scenario does not work as magnetic fields do not have time to leave the cloud through ambipolar diffusion. Therefore for supercritical clouds magnetic fields should be dragged into the star, forming stars with magnetizations far in excess of the observed ones (see Galli et al. 2006, Johns-Krull 2007). 

Magnetic fields are important at all stages of star formation. In many instances the ideas of star formation based exclusively on ambipolar diffusion have been challenged by observations (Troland \& Heiles 1986, Shu et al. 2006, Crutcher et al. 2009, 2010a, see Crutcher 2012 for a review).
While the interpretation of particular observations is the subject of scientific debates (see Mouschovias \& Tassis 2009), it is suggestive that there may be additional processes that the classical theory does not take into account. The primary suspect is turbulence, which is ubiquitous in interstellar media and molecular clouds\footnote{The presence of turbulence is an observational fact, while its sources are debatable. Supernova explosions, different
instabilities, e.g. magnetorotational instability, have been discussed in the literature. Flows induced by the gravitational instabiltiy (see Vazquez-Semadeni et al. 2011) are also
expected to be turbulent.} (see Larson 1981, Armstrong et al. 1994, Elmegreen \& Falgarone 1996, Lazarian \& Pogosyan 2000, Stanimirovic \& Lazarian 2001, Heyer \& Brunt 2004, Padoan et al. 2006, 2009, Chepurnov \& Lazarian 2010, Burkhart et al. 2010, 2012). Turbulence has revolutionized the field of star formation (see Vazquez-Semadeni et al. 1995, 2000, Ballesteros-Paredes et al. 1999, 2007, Elmegreen 2000, 2002, McKee \& Tan 2003, Elmegreen \& Scalo 2004, MacLow \& Klessen 2004, McKee \& Ostriker 2007), but the treatment of the turbulent magnetic fields stayed within the flux freezing paradigm.

The cornerstone concept of magnetic flux freezing has been challenged recently. On the basis of the model
of fast magnetic reconnection in Lazarian \& Vishniac (1999, henceforth LV99), Lazarian (2005) claimed that the removal of magnetic fields from turbulent plasma can happen during star formation due to magnetic reconnection rather than slow ambipolar drift (see also Lazarian \& Vishniac 2009). The process that was later termed {\it reconnection diffusion} does not depend on the degree of ionization, but rather on the properties of turbulence. The numerical confirmation of the idea was presented for molecular clouds and circumstellar accretion disks in Santos-Lima et al. (2010, 2012). The numerical testing of the LV99 reconnection model that is at the core of the concept of reconnection diffusion was successfully tested in Kowal (2009, 2012). Compared to this testing of reconnection diffusion, the testing in Kowal et al. (2009, 2012) was performed with much higher numerical resolution of the reconnection region and much better control of turbulence and other input parameters. In addition, more recent formal mathematical studies aimed at understanding of magnetic field dynamics in turbulent fluids supported the LV99 quantitative conclusions (Eyink 2011, Eyink, Lazarian \& Vishniac 2011). Moreover, the state of observations of magnetic fields in regions of star formation has progressed significantly (e.g. Crutcher 2012), so that comparisons of observational results with the theory are possible. These developments motivate us to further study the implications of the reconnection diffusion concept for star formation.

We note that the interstellar medium is collisional (Yamada et al. 2006), and therefore ideas of collisionless reconnection (see Shay \& Drake 1998, Shay et al. 1998, Bhattacharjee et al. 2005, Cassak et al. 2006) are not applicable. Even if they were, they would not increase the rate of 
reconnection diffusion, which is controlled by the eddy diffusivity. Similarly, the ideas related
to the plasmodial instability (see Shibata \& Tanuma 2001, Lourreiro et al. 2007, Uzdensky et al. 2010, Huang et al. 2011) do not change the reconnection rates either.\footnote{In special circumstances when
the initial configuration of magnetic field contains a magnetic reversal and is laminar, the plasmoid reconnection can induce turbulence in 3D (Karamabadi 2012). This may be important process for Solar flares (LV99, Lazarian \& Vishniac 2009, Eyink et al. 2011). However, observations 
(see Larson 1981, Armstrong et al. 1995, Chepurnov \& Lazarian 2010, Gaensler et al. 2011, see also Elmegreen \& Scalo 2006 and Lazarian 2009 for reviews) testify that interstellar media and molecular clouds are turbulent which makes LV99 mechanism directly relevant.}  

In what follows, in \S 2 we discuss the reconnection of magnetic fields in turbulent fluids, in \S 3 we discuss recent observational results reviewed by Crutcher (2012) that can be interpreted as contradicting the model of star formation mediated by ambipolar diffusion. We provide in \S 4 a very simple semi-analytical model of magnetic field diffusion out of magnetized clouds and discuss its predictions. In \S 5 we discuss other observational tests. Astrophysical implications of reconnection diffusion
are briefly discussed in \S 6, while the discussion and summary are provided in \S 7 and \S 8 respectively.

\section{Reconnection diffusion of magnetic field}

First of all, we should discuss how magnetic fields can diffuse in a conducting medium in the presence of
turbulence. For this purpose we should consider the process of magnetic reconnection. In the absence of
magnetic reconnection frozen-in magnetic field lines cannot cross each other, and the turbulent fluid should become increasingly entangled and eventually get locked. Using an analogy one can think of the fluid acquiring the properties of felt or Jello.

\begin{figure}
\centering
 \includegraphics[height=.30\textheight]{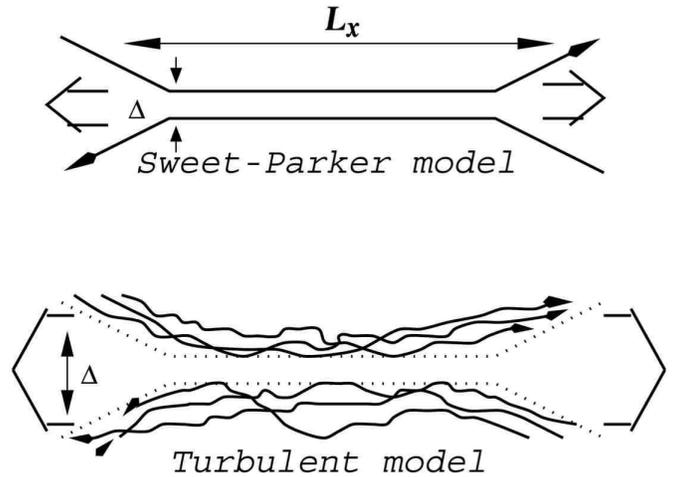}
 \caption{{\it Upper panel}. Sweet-Parker reconnection. $\Delta$ is limited by resistivity and is small. {\it Lower panel}: reconnection according to the LV99 model. $\Delta$ is determined by turbulent field wandering and can be large.  From Lazarian, Vishniac \& Cho (2004).}
\label{LV}
\end{figure}

According to LV99, magnetic reconnection of a turbulent field happens at a rate (see Appendix A)
\begin{equation}
v_{rec, LV99}\approx V_A (L_x/L)^{1/2}M_A^2 ~~~{\rm for}~L_x<L,
\label{LV99}
\end{equation}
where $L_x$ is the horizontal extent of the reconnection zone (see Figure \ref{LV}), $L$ is the injection
scale of turbulence, $V_A$ is the Alfv\'{e}n velocity and $M_A\equiv V_L/V_A$ is the magnetic Mach number. In
Eq. (\ref{LV99}) it is assumed that turbulence is subAlfv\'{e}nic. For $L_x>L$, LV99 got the expression that differs from Eq.~(\ref{LV99}) by a change of the power $1/2$ to $-1/2$. 

The enhancement of the reconnection rate compared to the Sweet-Parker model (see upper plot in Figure \ref{LV})
is achieved by increasing the outflow region $\Delta$ by accounting for magnetic field wandering. Indeed, in the Sweet-Parker model the reconnection speed is limited by the outflow of the matter from
the reconnection layer and $\Delta$ is determined by Ohmic diffusion. Therefore, 
\begin{equation}  
v_{rec}= V_A \frac{\Delta}{L_x} 
\label{vrec} 
\end{equation}
is much less than $V_A$  since $\Delta \ll L_x$. At the same time $\Delta$ becomes comparable with $L$
for $V\sim V_A$ provided that $\Delta$ is determined by the turbulent field wandering, which agrees well
with Eq. (\ref{LV99}).  

The rate given by Eq. (\ref{LV99}) is sufficient to allow the magnetic field to disentangle during the turnover of the eddies. Such eddies are a part of the picture of strong Alfv\'{e}nic turbulence (see Goldreich \& Sridhar 1995, henceforth GS95). The Alfv\'{e}nic cascade develops in compressible MHD turbulence
independently from the cascade of fast and slow modes, and its properties in compressible and incompressible
turbulence are very similar (Cho \& Lazarian 2002, 2003, Kowal \& Lazarian 2010). We use this cascade
in its GS95 description for the rest of the paper\footnote{Recent studies in Beresnyak (2011) and Beresnyak \& Lazarian (2010) indicate that the worries that the GS95 description may not present a perfect match for
numerical simulations are 
premature.}. For our arguments related to reconnection diffusion the compressible motions are of secondary
importance. Indeed, the LV99 reconnection model is governed by field wandering induced by the 
Alfv\'{e}nic cascade and turbulent mixing and also depends on the
solenoidal component of the fluid. 

The peculiarity of reconnection diffusion is that it requires nearly parallel magnetic field lines to reconnect, while
the textbook description of reconnection frequently deals with anti-parallel description of magnetic field lines.
One should understand that the situation shown in Figure \ref{LV} is just a cross section of the magnetic fluxes
depicting the anti-parallel components of magnetic field. Generically, in 3D reconnection configurations the sheared component of magnetic field is present. 

As we discuss below reconnection diffusion is closely connected with the reconnection between adjacent
Alfv\'{e}nic eddies (see Figure \ref{mix}). The GS95 model predicts that Alfv\'{e}nic turbulence is strongly anisotropic with eddies elongated along the local
direction of magnetic field. In other words, the extent of the eddy parallel to magnetic field $l_{\|}$
is larger than the extent in the perpendicular direction $l_{\bot}$. Mixing motions in MHD turbulence require that reconnection events in MHD turbulence should happen through every eddy turnover.  This is what the LV99 model predicts.  Indeed, for small scales magnetic field lines are nearly parallel and, when they intersect, the pressure gradient is not $V_A^2/l_{\|}$ but rather $(l_{\bot}^2/l_{\|}^3) V_A^2$, since only the energy of the component of the magnetic field that is not shared is available to drive the outflow. On the other hand, the characteristic length contraction of a given field line due to reconnection between adjacent eddies is $l_{\bot}^2/l_{\|}$. This gives an effective ejection rate of $V_A/l_{\|}$. Since the width of the diffusion layer over the length $l_{\|}$ is $l_{\bot}$, Eq.(\ref{LV99}) should be replaced by $V_{R}\approx V_A (l_{\bot}/l_{\|})$. This provides the reconnection rate $V_A/l_{\|}$, which is just the nonlinear cascade rate on the scale $l_{\|}$. This ensures self-consistency of critical balance for strong Alfv\'{e}nic turbulence in highly conducting fluids (LV99). 

\begin{figure}
  \includegraphics[width=0.95\columnwidth,height=0.24\textheight]{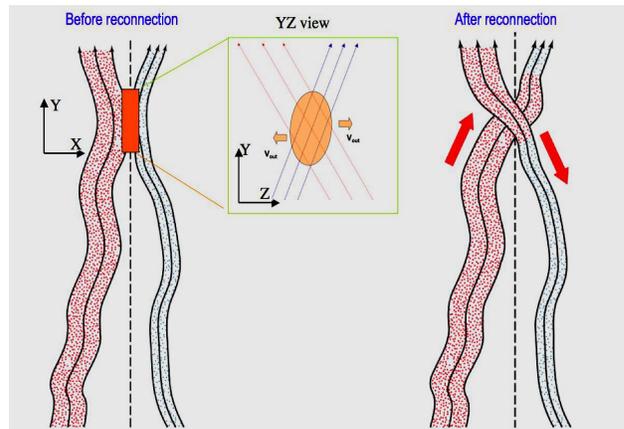}
  \caption{Motion of matter in the process of reconnection diffusion. 3D magnetic flux tubes get into contact and after reconnection plasma streams along magnetic field lines. {\it Left panel}: XY projection before reconnection, upper panel shows that the flux tubes are at angle in X-Z plane. {\it Right panel}: after reconnection. }
  \label{recdiff}
\end{figure}

Consider first a toy model illustrating how can plasmas move perpendicular to the mean inhomogeneous magnetic field (see Figure \ref{recdiff}). In the figure two magnetic flux tubes with entrained plasmas intersect each other at an angle and due to reconnection the identities of magnetic field lines change. The process of reconnection requires that the magnetic field lines are not parallel. The process is three
dimensional with magnetic field component reversing its direction in the Z-direction. Thus in XY plane
we observe nearly parallel magnetic fields, and the fact that magnetic field flux tubes cross each other at
an angle is seen in the subpanel of Figure \ref{recdiff}. Magnetic reconnection feeds upon free energy associated with the reversing magnetic field making magnetic fields smoother. For instance, it may be energetically favorable for magnetic flux tubes to reconnect, removing the Z-component of the reversed field
and create a smaller reverse component of the X-field (see the right part of the panel). The randomness
of the magnetic field in this case decreases, which well corresponds to the expectations of the dissipation
process. 

Let us assume that before the reconnection pressure of ionized gas $P_{gas}$ is larger in the left tube (red dots
in Figure \ref{recdiff}), but the total pressure $P_{gas}+P_{magn}$ is the same for the two tubes. In the
presence of reconnection, the flux tubes get connected and ionized gas streams from what used to  be the red
tube flows to the parts of the green tube. In other words, the process of reconnection changes the topology of the initial magnetic configuration and connects magnetic fields with different
 mass loading and gas pressures. As a result, ionized gas streams along newly formed magnetic field lines to equalize the pressure along flux tubes. Portions of magnetic flux tubes with higher magnetic pressure expand as plasma pressure increases due
 to the flow of plasma along magnetic field lines. The entropy of the system increases with magnetic and plasma pressures
 becoming equal within two newly formed flux tubes. This is the diffusion process mediated by reconnection
that we term ``reconnection diffusion''.  

Turbulence is a vital ingredient of reconnection diffusion. It both induces fast reconnection and transports
matter and magnetic fields via turbulent motions. Because of turbulence, in spite of action of reconnection
to smooth magnetic field, the magnetic fields cannot relax to minimal energy state. Tangled magnetic field
always emerges and magnetic reconnection inevitably follows.   
The process of entangling magnetic fields by eddies is illustrated by Figure \ref{mix} where it is shown how two adjacent turbulent eddies can induce magnetic field and matter diffusion. There while considering what is happening with two flux tubes we see how crossing flux tubes similar to those in Figure \ref{recdiff} naturally emerge all the time. The eddies of different sizes provide mixing of ionized gas at different 
scales, inducing the diffusion perpendicular to the mean magnetic field that is not limited by the speed
of ionized gas moving along magnetic field lines. 

Indeed, if the densities of plasma along magnetic field lines are different in the two flux tubes, the reconnection in Figure \ref{mix}
creates  new flux tubes with columns of entrained dense and rarefied ionized gas. The situation is similar
to that in Figure \ref{recdiff} and the gas moves along magnetic fields equalizing the
pressure within the newly formed flux tubes. As a result, eddies with initially different plasma pressure exchange matter and equalize plasma pressure. This process presents
the diffusion of ionized gas perpendicular to the mean magnetic field and the rate of this diffusion
does not depend on the gas ionization. The process happens at all scales of self-similar eddies constituting the turbulent cascade. The alternative microscopic picture of the
diffusion of magnetic field lines is provided in Appendix B. 

\begin{figure}
\includegraphics[width=0.95\columnwidth,height=0.24\textheight]{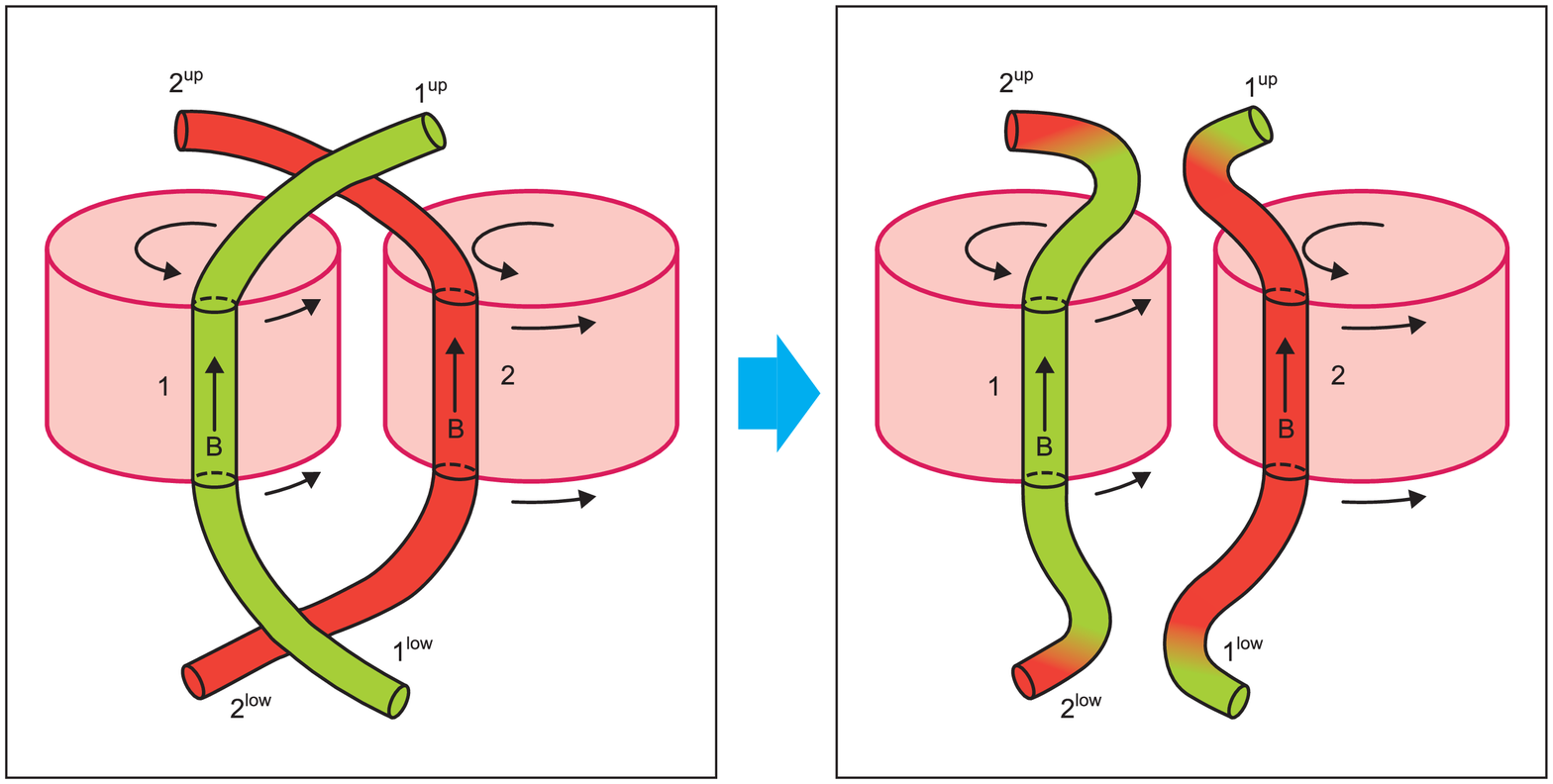}
\caption{Reconnection diffusion: exchange of flux with entrained matter. Illustration of the mixing of matter and
magnetic fields due to reconnection as two flux tubes of different eddies interact. Only one scale of turbulent
motions is shown. In real turbulent cascade such interactions proceed at every scale of turbulent motions.}
\label{mix}
\end{figure} 

In reality, for turbulence with the extended inertial range, the shredding of the columns of ionized gas with
different density proceeds at all turbulence scales making speed of plasma motion along magnetic field lines irrelevant for the diffusion. For the case of strong turbulence the diffusion of matter and magnetic
field over large scales $>L$ is given by Eq. (\ref{kappa_st}).  

Mixing motions of magnetic eddies induce turbulent diffusivity with the diffusion coefficient (Lazarian 2006):
\begin{equation}
\kappa_{strong}\sim L V_L M_A^3,
\label{kappa_st}
\end{equation}
where the reduction compared to the usual hydrodynamic turbulent diffusivity coefficient, i.e. 
\begin{equation}
\kappa_{hydro}\sim L V_L~~,
\label{hydro}
\end{equation}
arises from a relative inefficiency of mixing induced by weak turbulence. The latter is induced on the injection scale for isotropic subAlfv\'{e}nic turbulent driving, i.e. $M_A<1$. For superAlfv\'{e}nic turbulence the estimate given by Eq. (\ref{hydro}) is appropriate. 

Naturally, the estimate of diffusivity given by Eq. (\ref{kappa_st}) is applicable if the scales that
we deal with are larger than the injection scale of strong MHD turbulence. The latter is also
the scale for the transition from weak to strong turbulence (LV99, Lazarian 2006):
\begin{equation}
l_{trans}\sim L(V_L/V_A)^2\equiv LM_A^2.
\label{trans}
\end{equation}
If the scales that we deal with are smaller than the scale given by Eq. (\ref{trans}), then we deal with
the accelerating process of turbulent diffusion. Indeed, the separation between two particles $dl(t)/dt\sim v(l)$ for Kolmogorov turbulence is $\sim \alpha_t l^{1/3}$, where $\alpha$ is proportional to a cube-root of the energy cascading rate, i.e. $\alpha_t\approx V_L^3/L$ for turbulence injected with superAlvenic velocity
$V_L$ at the scale $L$. The solution of this equation is
\begin{equation} 
l(t)=[l_0^{2/3}+\alpha_t (t-t_0)]^{3/2},
\label{sol}
\end{equation}
which provides Richardson diffusion or $l^2\sim t^3$. The
accelerating character of the process is easy to understand physically. Indeed, the larger the separation 
between the particles, the faster the eddies that carry the particles apart. The character of the Richardson diffusion
in terms of field line separation\footnote{The description of the Richardson diffusion of magnetic field lines was provided in LV99.} is illustrated by the left panel of Figure \ref{regimes} in Appendix C.

When we deal with the turbulence within a cloud, it is plausible to associate the injection scale of the turbulence with the size of the cloud. Indeed, although the clouds may be a part of the compressible turbulence cascade originating at large scales, the disparity between the Alfv\'{e}n velocities 
in dense and rarefied gas mitigates the cross-talk between the eddies inside and outside the cloud. 

\section{Observations of magnetic fields in molecular clouds}

Star formation is known to be a rather inefficient process. Indeed, the interstellar mass of the Milky Way is $M_{MW}\approx10^9$ $M_\odot$. For the typical density of the gas of 50 cm$^{-3}$ the free fall time is $\tau_{ff}\approx
(3\pi/32G\rho)^{1/2}\approx 6\times 10^6$ years, which provides a ``natural'' star formation rate $M_{MW}/\tau_{ff}$ of 200 $M_\odot$ per year. At the same time the measured star formation rate is $\approx$ 1.3 $M_\odot$ per year (Murray \& Rahman 2010). It is because of this inefficiency that we still have large amounts ofÊÊ interstellar media in spiral galaxies. A traditional way of explaining this inefficiency is to appeal to 
magnetic forces preventing gravitational collapse (see Mouschovias \& Spitzer 1976). Indeed, it is possible to show that if the ratio of the magnetic flux to mass is larger than theÊ critical one,
\begin{equation}
(\Phi/M)_{crit}\approx 1.8\times 10^{-3} ~{\rm gauss}~{\rm cm}^2~g^{-1},
\label{crit}
\end{equation}
the magnetic field prevents cloud collapse (see Draine 2011) and star formation should come to a halt. If ambipolar diffusion mediates star formation, the
time scales are approximately right, which was a strong argument in favor of the ambipolar diffusion paradigm. 

A crucial parameter for star formation theory is therefore the ratio of mass to magnetic flux, $M/\Phi$, which parameterizes the relative importance of gravity and magnetic pressure. Strictly speaking, one should include {\em all} of the mass in a magnetic flux tube to calculate this ratio. However, what is important for star formation is the local ratio for a self-gravitating cloud. Other mass located far away in the flux tube will not affect the balance between gravity and magnetic pressure for the cloud. Observationally, one measures this local $M/\Phi$ by detecting the Zeeman effect in spectral lines of a species (such as OH or CN) that sample the densities and spatial regions of the cloud. Within a telescope beam of radius $r$ the magnetic flux being sampled is $\Phi \propto \pi r^2 B$, while the mass of the cloud within that beam is $M \propto \pi r^2 N$, where $N$ is the column density. Hence, the local $M/\Phi \propto N/B$. Since both $B$ and $N$ can be inferred from the Stokes $V$ and $I$ spectra, one infers the local $M/\Phi$; that is, the $M/\Phi$ in the region of space and velocity sampled by the observed species. This is the appropriate parameter for considering the balance between gravity and magnetic pressure in a cloud. This local $M/\Phi$ can change without magnetic diffusion if there are flows of matter onto (or from) the cloud from distant regions of the flux tube. Such flows provide a mechanism for building up a local self-gravitating mass (a cloud) without changing the local magnetic field strength. $M/\Phi$ could then vary from one place to another within a cloud. One issue is whether such flows can operate over the requisite distances without turbulence and instabilities disrupting the flow. A general discussion of such issues are beyond the scope of the present paper. However in Appendix D we provide analytical arguments showing that the accumulation along the magnetic field lines is unlikely in observations by 
Crutcher et al. (2011). Here we consider an alternative diffusion process to ambipolar diffusion for changing local $M/\Phi$.

The ambipolar diffusion theory of star formation is well developed, with several specific predictions. Crutcher (2012) reviewed the current state of observations of magnetic fields in molecular clouds and compared the observations with the predictions. Here we briefly summarize four observational tests of the theory.

\begin{figure}
\centering
Ê\includegraphics[height=.27\textheight]{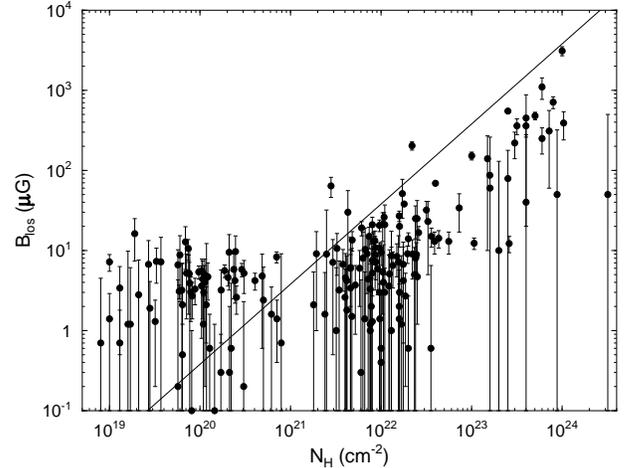}
Ê\caption{ HI, OH and CN Zeeman measurements of line of sight component of magnetic field versus the total column density $N_H$Ê of atomic and molecular hydrogen. The straight line is the critical mass/flux $M/\Phi =3.8\times N_H/B$ with subcritical clouds above the line. The two molecular clouds near $N ~ 10^{22}$ have OH maser contamination in their spectra and the corresponding Zeeman results are not reliable. From Crutcher 2012.}
\label{N}
\end{figure} 

First, ambipolar diffusion predicts and requires that $M/\Phi$ increase from the envelope to core region of a cloud. Crutcher, Hakobian \& Troland (2009) directly tested this prediction. Their experiment measured ${\cal R} \equiv (M/\Phi)_{core}/(M/\Phi)_{envelope}$. The idealized ambipolar diffusion theory of core formation requires ${\cal R}$ to be approximately equal to the inverse of the original (before evolution driven by ambipolar diffusion) subcritical $M/\Phi$, or ${\cal R} > 1$. The test used OH Zeeman observations to determine the line-of-sight magnetic field $B_{los}$ and the column density $N_{OH}$ in each of four dark cloud cores and envelopes. The central assumption of this test was that the angle $\theta$ between the line of sight and the magnetic field was approximately the same in core and envelope. Then the measured $B_{los} = B_{tot} \cos\theta$ would have the same $\theta$ in the numerator and denominator of ${\cal R}$, and the dependence on the unknown $\theta$ would disappear.Ê Therefore, ${\cal R} = (N_{OH}/B_{los})_{core}/(N_{OH}/B_{los})_{envelope}$, all directly measurable quantities. 

Because cloud cores with the strongest measured $B_{los}$ were selected for the test, it is likely that $B_{los} \approx B_{tot}$ with $\theta \sim 0$, so $\cos\theta \approx 1$ and small changes in $\theta$ between core and envelope would not affect ${\cal R}$ very much. Supporting this assumption of the experiment is the fact that published models of core formation driven by ambipolar diffusion have strong, regular magnetic field morphology such that the unknown angle $\theta$ between {\bf B} and the line of sight is approximately the same in core and envelope regions. Observationally, field morphologies in cores and envelopes are found to be correlated in direction (Li et al. 2009). 

With measurements of $B_{los}$ and $N(OH)$ toward each core and envelope, it was possible to calculate ${\cal R}$ and its uncertainty toward each cloud. Because $B_{los}$ was not detected (at the $3\sigma$ level) toward any of the four cloud envelopes, in no case was the calculated value of ${\cal R}$ significantly ($3\sigma$) different from 0. But the astrophysical question is not whether ${\cal R}$ is significantly different from 0, it is whether the ${\cal R}$ were significantly different from 1, since the ambipolar diffusion theory requires that ${\cal R} > 1$. The results for the four clouds were: ${\cal R}(L1448CO) = 0.02 \pm 0.36, {\cal R}(B217-2) = 0.15 \pm 0.43, {\cal R}(L1544) = 0.42 \pm 0.46$, and ${\cal R}(B1) = 0.41 \pm 0.20$. Hence, in all four cases, the experiment found ${\cal R} < 1$, not ${\cal R}>1$ as the theory predicts. In two of the four cases, ${\cal R} < 1$ at near a $3\sigma$ level. 

Proponents of the ambipolar diffusion theory of star formation strongly attacked the Crutcher, Hakobian \& Troland (2009) results and conclusions (Mouschovias \& Tassis 2009, Mouschovias \& Tassis 2010). The four strongest objections and the counters to those objections are addressed below. 

(1) Envelope regions around the core positions are not uniform, so the four envelope results should not be combined to improve the signal-to-noise ratio. For each envelope position considered separately, the sensitivity to the differential $M/\Phi$ is reduced and the disagreement with ambipolar diffusion less significant. However, at all four envelope positions in all four clouds, the column density is smaller than that at each central position, so envelope positions are being measured, just not the identical envelope column density at each position. The mean envelope column density is about 1/2 that of each core position. Hence, by obtained a mean $M/\Phi$ for each envelope, one is measuring mean envelope properties that are entirely suitable for comparison with the central core properties for each cloud. This procedure is similar to the stacking operation used in galaxy studies to infer mean properties around the center of galaxies, and is an entirely accepted and appropriate technique for obtained mean properties. 

(2) Since $B_{los}$ was not detected in the envelopes only upper limits should be considered. However, using the obtained value and its uncertainty to compare with the prediction of theory is equally valid. If theory predicts 1 and observations give $0.5 \pm 0.25$, it is valid to state that the experiment differs from the prediction by $2\sigma$, which is the approach we take. Their preference for stating that the data {\em agree} with the theory at the $2\sigma$ level is not a different statement. In no case do Crutcher, Hakobian \& Troland (2009)Ê claim a result $>3\sigma$, but merely state that the data for all four clouds do not correspond to the result predicted by theory. Moreover, the assumption of the theory that clouds initially are subcritical implies not that ${\cal R} =1$ but that ${\cal R} > 1$, so the significance of the individual observational results for ambipolar diffusion is larger than given by comparison with ${\cal R} = 1$. 

(3) TheÊ motion of cores through surrounding more diffuse gas could lead to {\bf B} in cores and their envelopes not being essentially parallel. However, while this is certainly possible, it would require that all four cores move nearly in the plane of the sky so the envelope fields would be dragged into the plane of the sky. Also, as mentioned above, evidence from linear polarization mapping that directions in cores and surrounding gas are strongly correlated (Li et al. 2009) argues against this morphology. 

(4) Multiplication of the probabilities of the data for each cloud being consistent with ambipolar diffusion to assess agreement with theory is invalid because the four clouds are measurements of different cases, so probabilities cannot be combined. However, while the observational results for each cloud individually can to argued to be (marginally) consistent with the prediction of ambipolar diffusion, one must also consider that a sample of four clouds was observed. If in fact ${\cal R} > 1$ in all four clouds, one would expect observational noise to sometimes produce an observed ${\cal R}$ greater than the ``true'' value, not always smaller. While the sample size of four clouds is not large, the observations reported by Crutcher, Hakobian \& Troland (2009) certainly suggest a problem with the idea that ambipolar diffusion was responsible for the formation of cores in {\em all four} clouds. If the probability for one cloud is $p_1$, it is correct to calculate $p_1p_2p_3p_4$ as the probability that {\em all four} were formed by ambipolar diffusion. Hence, the probability stated by Crutcher, Hakobian \& Troland (2009) that {\em all four} of the clouds have ${\cal R} > 1$ is $3 \times 10^{-7}$ is statistically correct.Ê 

Second, although the $M/\Phi$ ratio is an input parameter to the ambipolar diffusion theory of star formation, for theÊ theory to be relevant clouds must initially be subcritical, with ambipolar diffusion leading to an increase of mass in cores until cores become supercritical and collapse. The physics of the theory would be perfectly valid if subcritical clouds did not generally exist in nature, but then ambipolar diffusion would not be the physics driving star formation. Zeeman data make it possible to assess whether relatively low-density self-gravitating molecular clouds are generally subcritical. Figure~\ref{N} from Crutcher (2012) shows Zeeman observations of $B_{los}$ versus $N_H$. Although an individual Zeeman measurement gives $B_{los}$ (hence only an upper limit to $M/\Phi$), the upper envelope of the $B_{los}$ in Figure~\ref{N} defines the total field strength $B_{tot}$ at each $N_H$, since for some fraction of the clouds {\bf B} should point approximately along the line of sight. For $N_H \lesssim 10^{21}$ cm$^{-2}$, the $M/\Phi$ are subcritical; these clouds are lower density H~I clouds. For $N_H \gtrsim 10^{21}$ cm$^{-2}$, the $M/\Phi$ are overwhelmingly supercritical; these clouds are higher density molecular clouds and cores. Hence, the data appear consistent with the theory, with neutrals gravitationally contracting while leaving the magnetic flux behind and hence increasing $M/\Phi$ in the higher density molecular gas. However, the H~I clouds are in approximate pressure equilibrium with the warm interstellar medium and are not self-gravitating, so they could not gravitationally collapse through the magnetic field to form supercritical cores. For $N_H \gtrsim 10^{21}$ cm$^{-2}$ most of the points in Figure~\ref{N} are molecular clouds that are generally self-gravitating, so this should be the region of transition from subcritical to supercritical clouds. Yet there are zero definite cases of subcritical clouds for $N_H > 10^{21}$ cm$^{-2}$! The two points which seem to be significantly above the critical line are both from OH in absorption toward H~II regions, with polarized maser emission within the absorption lines producing inconsistencies and uncertainties in the fit for $B_{los}$ that were not reflected in the formal uncertainties. Hence, there are no clear observational cases of self-gravitating, subcritical clouds, which must exist in the ambipolar diffusion theory of star formation.

\begin{figure}
\centering
\includegraphics[width=\columnwidth]{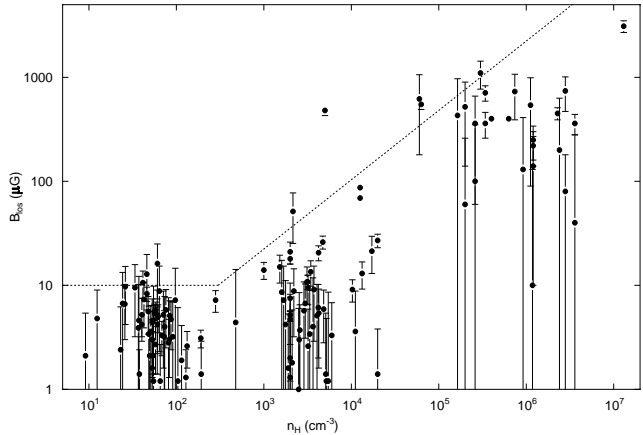}
\caption{H~I, OH, and CN Zeeman measurements of the magnitude of $B_{los}$ versus $n_H$. The dashed line segments show the most probable model from the Bayesian analysis for the maximum total magnetic field strength at each density. The total field strength of an individual cloud lies between this maximum and essentially 0, with a flat distribution between 0 and the maximum; from Crutcher et al. (2010b).}
\label{fig:zeeman}
\end{figure}

The third and fourth tests come from a statistical analysis of Zeeman data. For ambipolar diffusion to be the principal driver of star formation, magnetic field strengths in molecular clouds should be strong, so that initially clouds are subcritical. Ambipolar diffusion does not decrease field strengths, it has neutrals collapsing through magnetic fields essentially leaving the field in place until the core becomes supercritical. The field strength in the core would then increase with density. For a $B \propto \rho^k$ scaling parameterization, $k$ would increase from near 0 to 0.5 as the collapse proceeded. Hence, one would expect the total magnetic field strengths in molecular clouds to be high so $M/\Phi$ would be near critical, and the field to be sufficiently strong that it controls the power law exponent $k$ to be within the above range. Crutcher et al. (2010) carried out a Bayesian statistical analysis of Zeeman data in order to infer the PDF of $B_{tot}$ from the observed $B_{los}$ and compared results with these final two predictions of the ambipolar diffusion theory. Figure~\ref{fig:zeeman} shows the Zeeman measurements of $B_{los}$ versus $n_H$ used in the Bayesian analysis. 

The third prediction of ambipolar diffusion is that most molecular clouds should be subcritical, except for those that have achieved supercritical cores and are collapsing; no very weak magnetic fields should be found. The most probable PDF of $B_{tot}$ was found to be a flat function; that is, for a set of clouds at any density $n_H$ the $B_{tot}$ were uniformly distributed between very small values and the maximum $B_{tot}$ found for that $n_H$ (the dashed line shown in Figure~\ref{fig:zeeman}. The mean $M/\Phi$ for molecular clouds was found to be supercritical by $\sim3$. Hence, as noted above, {\em no} subcritical molecular clouds have been shown to exist, so all $M/\Phi >1$ (supercritical), and many molecular clouds have very small $B_{tot}$ and hence are {\em very} supercritical. 

Fourth, the Bayesian analysis found that $k \approx 0.65 \pm 0.05$ for the $B \propto \rho^k$ scaling of total field strength with density. The ambipolar diffusion theory predicts $\kappa < 0.5$. A $k = 2/3$ value would be produced by contracting clouds with weak magnetic fields, such that magnetic pressure would not significantly affect the geometry of the collapse (Mestel 1966). Such weak fields are consistent with the result for the PDF of total field strengths just discussed.

There are thus four discrepancies between the predictions of ambipolar diffusion theory and observations of magnetic fields in regions of star formation for the existing paradigm of star formation -- near flux freezing with ambipolar diffusion driving contraction of cloud cores. In the next two sections we will consider whether reconnection diffusion theory is consistent with these four observational results.

\section{Simple model for reconnection diffusion}

\begin{figure*}
\center
\includegraphics[width=0.9\columnwidth]{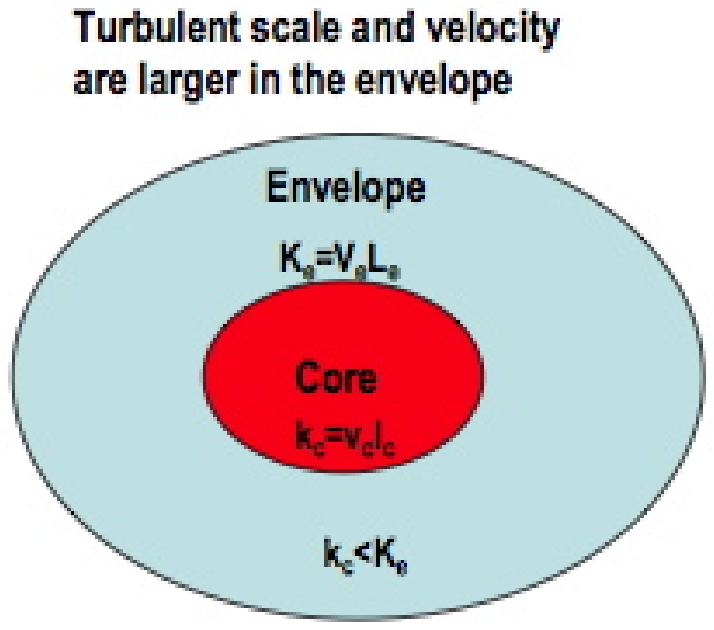}
\includegraphics[width=0.9\columnwidth]{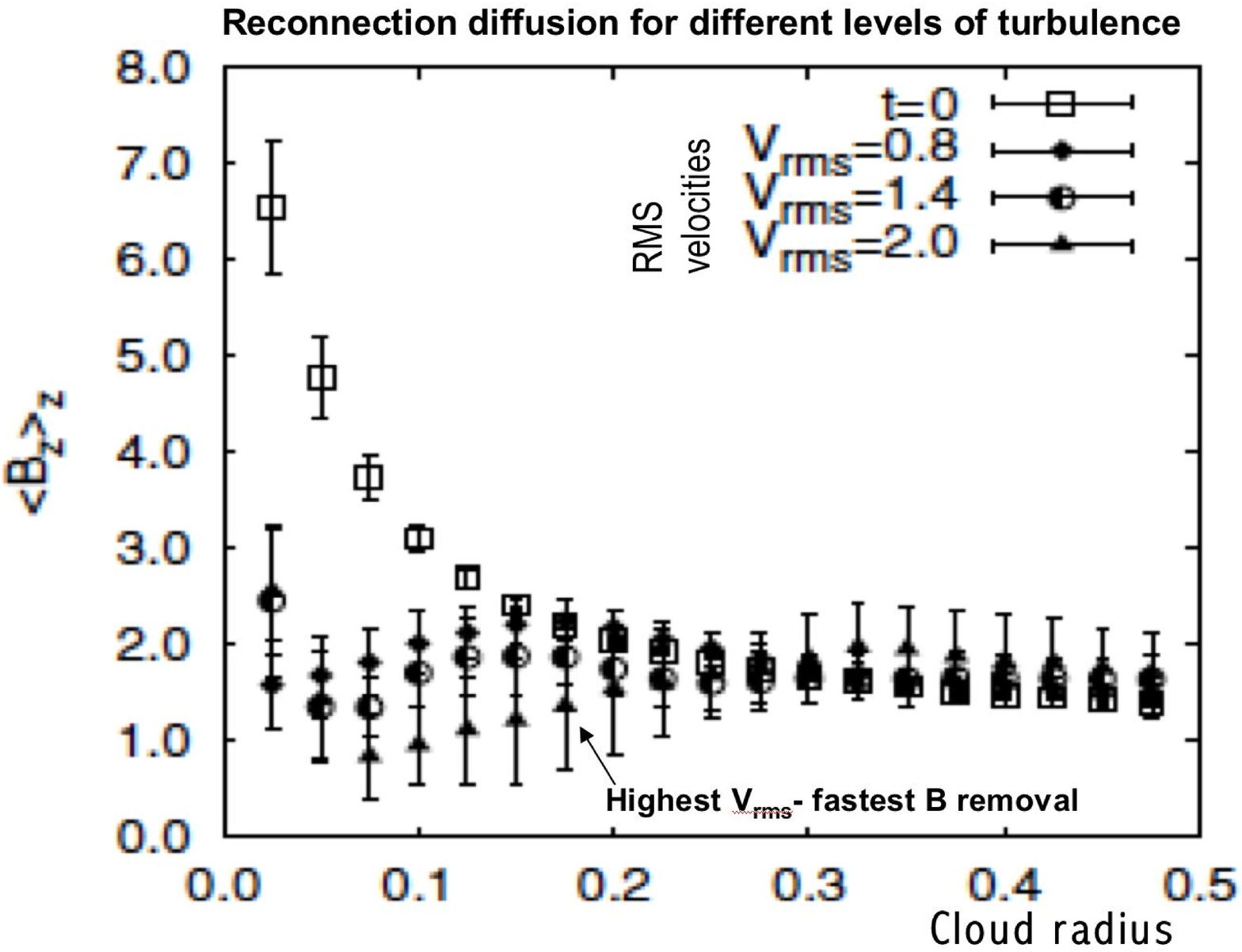}
\caption{{\it Left panel}. Schematic of the cloud core and envelope in Crutcher et al. (2009). Both scale of the turbulence and velocity (and thus reconnection diffusion rate) are larger in the envelope than in the core. {\it Right panel}. Results of 3D MHD
simulations (modified from Santos-Lima et al. 2010) The removal of magnetic field via reconnection diffusion increases with the increase of the turbulence level.
\label{crutcher}}
\end{figure*}

We first consider the first test of the above section. We will directly compare a reconnection diffusion model with the results of the Crutcher, Hakobian \& Troland (2009) observations that $M/\Phi$ {\em decreases} rather than {\em increases} between envelopes and cores of the four clouds they studied. The relevant mean parameters for the four cores are $B_{total} \approx 30$ $\mu$G, $n_H \sim 10^4$ cm$^{-3}$, and radius $\approx 0.2$ pc, and for the envelopes $B_{total} \sim 5-10$ $\mu$G, $n_H \sim 10^3$ cm$^{-3}$, and radius $\sim 1$ pc. If these clouds formed by contraction of a typical diffuse H~I cloud, the parameters would be $B_{total} \approx 5$ $\mu$G, $n_H \sim 60$ cm$^{-3}$, and radius $\approx 2.5$ pc.

As we mentioned earlier, in the presence of gravity reconnection diffusion allows density to concentrate towards the gravitational center, while magnetic field leaves the gravitational potential. The reconnection diffusion rate and thus the rate of magnetic field removal is expected to be proportional to the level of turbulence.
Simulations that illustrate this effect are presented in Figure~\ref{crutcher}. It is known that the velocities
in cloud envelopes are larger than the velocities in the cloud cores (see Taffalla et al. 1998). In addition, the scale
of turbulence involved in reconnection diffusion is also larger. Thus we expect the diffusion coefficient for the envelope (see Eq. \ref{kappa_st}) to be larger compared to the core coefficient, and it is
natural that faster removal of magnetic field happens from the envelope than from the core. In fact, we
also expect the uniformly compressed magnetic field to be smaller at the edges of the cloud even
for uniform diffusivity.  
These theoretical expectations agree with the results in Crutcher,
Hakobian \& Troland (2009).
The observational results disagree with the ambipolar diffusion paradigm as the rate of ambipolar diffusion decreases towards edges of clouds, and within clouds the ambipolar diffusion of magnetic fields is negligible for media with high degree of ionization. On the contrary, reconnection diffusion does not depend
on the degree of ionization, and it can successfully proceed in highly ionized gas. 

In what follows, we illustrate the effects of reconnection diffusion
using simplified calculations of magnetic diffusivity for conditions motivated by the observations in Crutcher et al. (2009).

\subsection{Solutions of the model: the radial profile produced by magnetic reconnection}
Magnetic fields are weightless. Therefore we can get an insight in the physics of
the reconnection diffusion by modeling the diffusion of magnetic field from the place of
its original concentration. For illustrative purposes we shall limit our discussion to
the one dimensional diffusion:
\begin{equation}
\frac{\partial B}{\partial t}-\kappa\frac{\partial^2B}{\partial x^2}=0.
\label{eq:diff}
\end{equation}
This equation can be solved analytically for a constant diffusion coefficient
$\kappa$ and an initial distribution of magnetic field
\begin{equation}
B(x,t=0)=\phi (x).
\label{inial}
\end{equation}
The solution of the equation is given by
\begin{equation}
B(x,t)=\int^{\infty}_{-\infty} \frac{\phi\left(x' \right)}{2\sqrt{\pi\,\kappa\,t}}
\,\mathrm{e}^{-\frac{(x-x')^2}{4\kappa t}}\,\mathrm{d}x',
\label{sol}
\end{equation}
which can be easily solved for the case of a uniformly compressed fluid
\begin{equation}
\phi(x)=\cases{
 B_{0}, & for $\vert x\vert \le R$\cr
 B_{\mathrm{ISM}}, & for $\vert x\vert>R$,
}
\label{initcond}
\end{equation}
where $ B_{0}$ denotes the initial value of the magnetic field after
the compression, and  $B_{\mathrm{ISM}}$ the magnitude of the field
outside.
The solution can be obtained considering a constant boundary
condition  (i.e. $B(x,t)=B_{\mathrm{ISM}}$, for $\vert x \vert > R$), and
it is given by the error function
\begin{eqnarray}
B(x,t)&=&B_{\mathrm{ISM}}+\left(B_0-B_{\mathrm{ISM}}\right)\times\nonumber \\ &&\left[\mathrm{Erf}\left(\frac{R-x}{2\sqrt{\kappa\,t}} \right)
                      +\mathrm{Erf}\left(\frac{R+x}{2\sqrt{\kappa\,t}} \right) \right].
\label{sol2}
\end{eqnarray}

Let us estimate the reconnection diffusion coefficients based on the
observations of Crutcher et al. (2009). 
We have considered a cloud core with size (radius)
$\ell=0.2~\mathrm{pc}$, within an envelope of radius
$L=1.0~\mathrm{pc}$. 
For a magnetic field of $B=10$ $\mu\mathrm{G}$ in the envelope, and a
hydrogen number density 
$n_H=10^3\,\mathrm{cm}^{-3}$, the Alfv\'{e}n speed is
\begin{equation}
V_{A,\mathrm{env}}=\frac{B_{\mathrm{env}}}{\sqrt{4\pi\,n_{H,\mathrm{env}}\,\mu\,m_H}}\approx 0.6~\mathrm{km~s^{-1}}
\end{equation}
for $\mu=1.3$ (neutral gas with solar abundances). 
For the core we assumed $B=35$ $\mu\mathrm{G}$, and
$n_H=10^4\,\mathrm{cm^{-3}}$, which results in a similar $ V_{A,\mathrm{core}} \approx
0.67~\mathrm{km~s^{-1}}$. 

For the velocity we use the mean dispersion obtained from
the OH lines in the clouds observed by Crutcher et al. (2009). They
measure radial velocities with a mean full width at half maximum of
0.99 $\mathrm{km~s^{-1}}$ and 0.76 $\mathrm{km~s^{-1}}$, for the
envelope and core, respectively. These correspond (dividing by
$2\sqrt{2\ln 2}\approx 2.35$) to a velocity dispersion of
0.42 $\mathrm{km~s^{-1}}$ and 0.32 $\mathrm{km~s^{-1}}$, respectively.
Assuming that the motions are isotropic, the magnitude of
the total velocity dispersion can be obtained by multiplying those
values by $\sqrt{3}$,  yielding $V_L\approx 0.73~\mathrm{km~s^{-1}}$
in the envelope, and  $V_{\ell}=0.55~\mathrm{km~s^{-1}}$ in the core.

The resulting Alfv\'{e}nic Mach numbers are $M_{A,\mathrm{env}}\sim
1.2$ for the envelope, which falls into the
strong turbulence regime, and $M_{A,\mathrm{core}}\sim 0.8$,
corresponding to the weak turbulence case. 

Let us illustrate the diffusion of magnetic field due to reconnection
by taking two discrete values for the $\kappa$ coefficient, one for the core
$\kappa_{\mathrm{core}}$  (for weak turbulence) and one for its envelope
$\kappa_{\mathrm{env}}$ (for strong turbulence), such that:
\begin{eqnarray}
\kappa_{\mathrm{core}} = & V_{\ell}\,\ell\,M_{A,\mathrm{core}}^3 &
\approx 1.86\times 10^{22}~\mathrm{cm^2~s^{-1}}, \\
\kappa_{\mathrm{env}}  = & V_{L}\,L & \approx 2.25\times 10^{23}~\mathrm{cm^2~s^{-1}}.
\end{eqnarray}
Which gives  a ratio of $\kappa_{\mathrm{env}}/\kappa_{\mathrm{core}}
\sim 12$; thus the rate of removal of magnetic field through turbulent
reconnection is roughly twelve times faster with $\kappa_{\mathrm{env}}$,
with a diffusion timescale of $t_{\mathrm{diff,env}}=\frac{L^2}{\kappa_{\mathrm{env}}}\approx
1.3~\mathrm{Myr}$.
\begin{figure}
\centering
\includegraphics[width=\columnwidth]{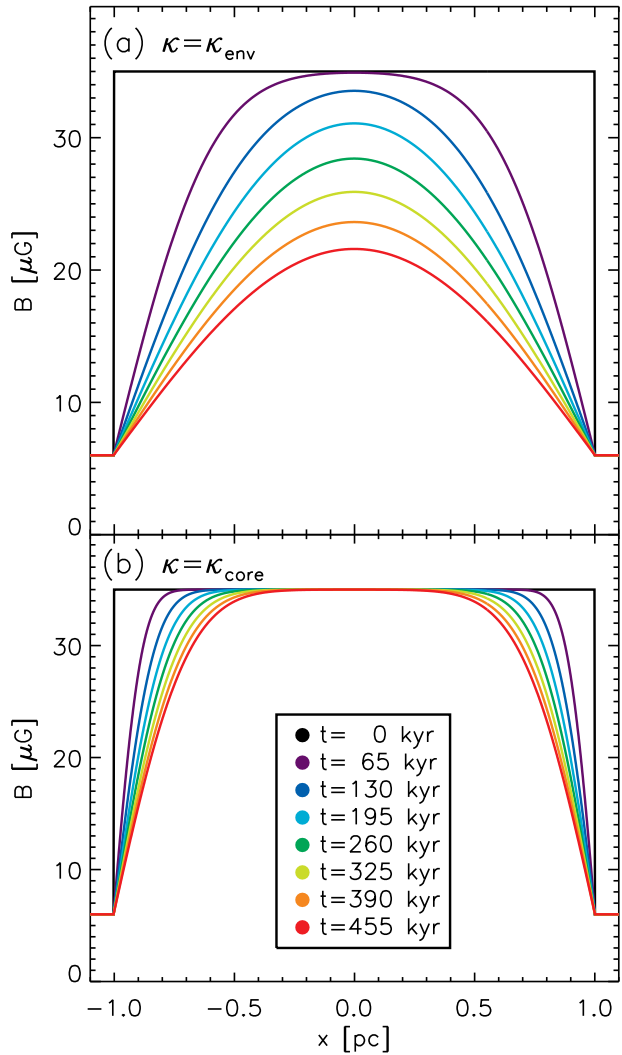}
\caption{Diffussion of the magnetic field with a constant diffusion coefficient
(Equation \ref{sol2}). In the top panel (a) the solution is obtained
with the turbulent diffussion coefficient $\kappa_{\mathrm{env}}$, and
in the bottom panel (b) the same solution, but with the coefficient in
the core ($\kappa_{\mathrm{core}}$) is presented. 
The different lines, as indicated by the color coding in the legend,
correspond to different times at intervals of $65$~kyr, from $t=0$ to
$t=455$ kyr.} 
\label{fig:kconst}
\end{figure}
In Figure \ref{fig:kconst} we illustrate the solution of Eq.
(\ref{sol2}) for the initial  
conditions given in Eq. (\ref{initcond}), with $R=L=1.0~\mathrm{pc}$,
$B_0=B_{\mathrm{core}}=35~\mu\mathrm{G}$, and
$B_{\mathrm{ISM}}=6~\mu\mathrm{G}$. In the figure we see how the
magnetic field diffuses from the envelope towards the core, producing
already a profile that is consistent with the observations (i.e. a larger
field in the core than in the envelope). Of course, this is an overly
simplified model, as the diffusion coefficient is not constant
throughout the region. It gives, however, an insight of the timescales
involved in either case.

We also search numerically for a solution assuming that the diffusion
coefficient $\kappa$ scales as $l^{\alpha}$, $\alpha>0$, which
represents the increase of the diffusion efficiency with the scales,
which is expected from turbulent diffusivities. We assume for
simplicity that $\alpha=4/3$ corresponding to the GS95 picture of turbulence.
The scale dependence is $\kappa \propto \ell^{4/3}$, 
where the proportionality constant was chosen so that
$\kappa=\kappa_{\mathrm{core}}$ for $x\rightarrow 0$, and
$\kappa=\kappa_{\mathrm{env}}$ for $x=L$. 

\begin{figure}
\centering
\includegraphics[width=\columnwidth]{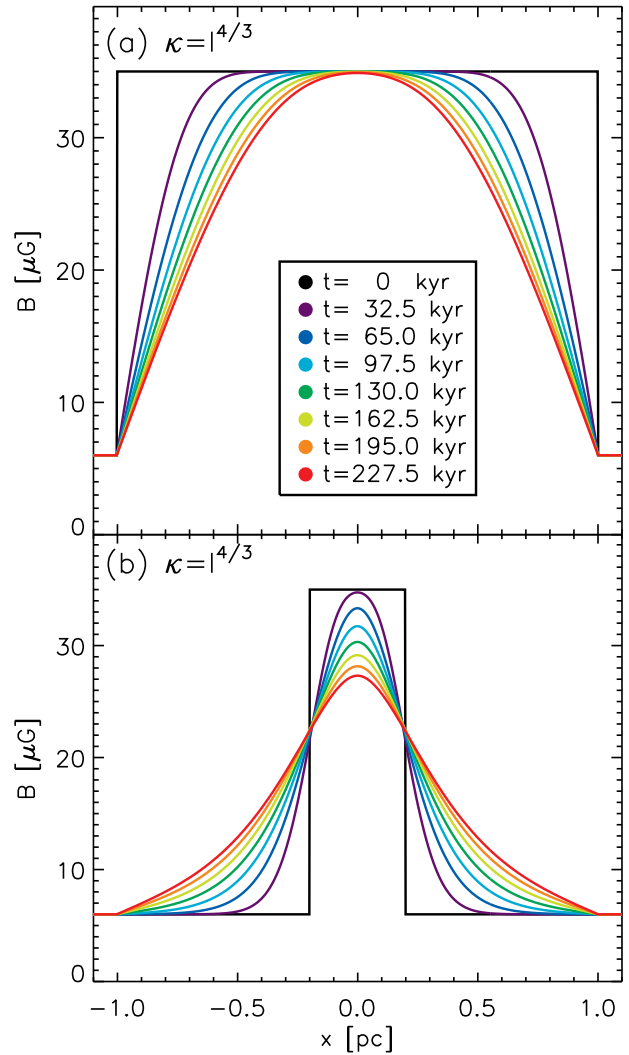}
\caption{Diffusion of the magnetic field with a scale dependent diffusion coefficient,
 $\kappa \propto \ell^{4/3}$. 
The initial condition for the magnetic field is represented by the
black line, a single step in panel (a), and a two step in panel (b),
see text for details.
The different lines, as indicated by the
 color coding in the legend, correspond to different times at
 intervals of $32.5$ kyr, from $t=0$ to $t=227.5$ kyr.} 
\label{fig:kvar}
\end{figure}

The solutions (numerical) with a scale dependent coefficient are presented in Figure
 \ref{fig:kvar} for two different initial conditions. In the top
 panel, the same field used in Figure \ref{fig:kconst} is used. In the
 bottom panel we have used:
\begin{equation}
\phi(x)=\cases{
 35~\mu\mathrm{G}, & for $\vert x\vert \le \ell $\cr
 ~6~\mu\mathrm{G}, & for $\vert x \vert > \ell. $
}
\label{initcond2}
\end{equation}
In this solutions we can see the effect of a more efficient removal of
magnetic field in the envelope than in the core. In the case in which
the initial field was concentrated at the core (Figure
\ref{fig:kvar}b) it is also evident how some of the field from the
core diffuses to the envelope. In that case and after $\sim 200$ kyr,
the magnitude, both for the core and the envelope are comparable to
the observations of Crutcher et al. (2009).
An important point is that, regardless of the initial condition, the
profiles preserve an increasing magnetic field towards the center,
with a larger strength ratio between center and envelope than ambipolar diffusion would produce.

\section{Other observational tests}

\subsection{No subcritical molecular clouds}

The second test discussed in \S3 is the fact that there is no observational evidence for the subcritical self-gravitating molecular clouds predicted by the ambipolar diffusion theory to be observable among the sample of several dozen dark clouds (densities $10^{3-4}$ cm$^{-3}$) for which OH Zeeman observations are available. Such clouds are the initial condition for the ambipolar diffusion theory to be relevant to star formation. On the other hand, this observation is a natural consequence of reconnection diffusion, which acts in low-density clouds where turbulence is strong to prevent an increase of magnetic field strength with density. In low-density clouds the kinetic and magnetic energies are approximately in equipartition and larger than the thermal energy (Heiles \& Troland 2005). From the point of view of our analysis, this is a signature of the transAlfv\'{e}nic supersonic turbulence indicating that the turbulence should be efficient in driving reconnection diffusion. We note, however, that an alternative picture for building up self-gravitating clouds without significantly increasing magnetic field strengths is the collection of matter along magnetic field lines. In a simplistic picture of this process, density would increase while the magnetic field strength would be unchanged. Such a process would produce thin cylinder or disk-like clouds with the magnetic field along the minor axis. However, there is not universal observation support for this morphology. Goodman et al. (1990) and Chapman et al. (2011) found clouds with minor axes parallel, perpendicular, and intermediate angles with respect to the surrounding magnetic field. A statistical analysis by Tassis et al. (2009) found that the magnetic fields were oriented close to ($\sim 24^\circ$) the minor axis but could not reject alternative configurations. No doubt matter can accumulate more along than perpendicular to magnetic fields. But the scales for such one-dimensional collection to form GMCs are of hundreds of parsecs (see Vazquez-Semadeni et al. 2011, Appendix D.). Field wandering induced by turbulence (see Appendix D) would tend to interfere with a smooth flow over such distances. We believe that reconnection diffusion is an intrinsic part of the simulations of the turbulent interstellar medium, and it should be accounted for in interpreting the results of turbulent numerical simulations.

\subsection{Clouds with very weak fields}

The third observational result that ambipolar diffusion theory has difficulty with is the flat PDF of total field strengths (Crutcher et al. 2010); that is, at a given density field strengths range from very small values corresponding to a highly supercritical $M/\Phi$ up to values that correspond to an approximately critical $M/\Phi$. The molecular clouds with very small field strengths could arise from the action of reconnection diffusion over time, with older clouds having field strengths approaching that of the surrounding H~I medium. While ambipolar diffusion theory does not prohibit the observed highly supercritical molecular clouds at relatively low densities ($10^{3-4}$ cm$^{-3}$), for the theory to be relevant most clouds must start as subcritical. 

\subsection{Scaling of field strength with density}

The fourth observational result, that field strengths scale as density to the 2/3rds power at high densities (Figure \ref{fig:zeeman}), was predicted by Mestel (1966) for contracting clouds with magnetic fields too weak to significantly affect the collapse. We believe that the necessary low field strengths are due to reconnection diffusion being efficient at the lower densities before collapse. Once rapid collapse begins, reconnection diffusion removes magnetic field from the contracting clouds more slowly than contraction increases it. If the reconnection diffusion time 
\begin{equation}
t_{rec. diff}=l^2/\kappa~~,
\label{rec}
\end{equation}
where $l$ is the cloud size, is less than the free fall time 
 \begin{equation}
 t_{ff}=\sqrt{3\pi/(32 G \rho)}~~,
 \label{ff}
 \end{equation}
reconnection diffusion should be efficient in keeping the magnetic field inside the cloud close to the value of the magnetic field outside 
the cloud (see discussion in \S 5.2).  
 
 If this is true, for the relevant scales reconnection diffusion must be faster than free fall. 
Consider the scales at which $t_{rec, diff}$ (see Eq. (\ref{rec})) is less than the free fall time $t_{ff}$ (see Eq. (\ref{ff})) 
and compare our expectations with observations. 
If for the case of subAlfv\'{e}nic turbulence, one uses the scale $l$ of the cloud and the corresponding velocity 
\begin{equation}
v_l=V_{inj} (l/L_{inj})^{1/3},
\label{Kolm}
\end{equation}
where  $L_{inj}$ and $V_{inj}$ are the injection scale and velocity, respectively, then the Kolmogorov scaling of velocities given by Eq. (\ref{Kolm}) corresponds to the predictions of the GS95 model for scales less than $l_{trans}=L_{inj} (V_L/V_A)^3$ (see Lazarian 2006). Thus one can get the condition that reconnection diffusion is faster than cloud free fall, i.e. $t_{rec. diff}<t_{ff}$, if
\begin{equation}
l_{lower}>\left(\frac{32}{3\pi}\right)^{3/2} L_{inj} \left(\frac{L_{inj} (G\rho)^{1/2}}{V_{inj}}\right)^3 \left(\frac{V_A}{V_{inj}}\right)^9~~.
\label{case1}
\end{equation}
In the case that magnetic field stays roughly constant as in Figure \ref{fig:zeeman}, the Alfv\'{e}n velocity $V_A$ in Eq. (\ref{case1}) changes with density as $\rho^{-1/2}$, and
the requirement on the scale of the cloud can be rewritten
\begin{equation}
l_{lower}>\frac{1}{\pi^6 6^{3/2}}\frac{L_{inj}^4}{V_{inj}^{12}} \frac{G^{3/2} B^9}{\rho^3}~~,
\label{case1'}
\end{equation}
which provides the lower constraint on the size of the cloud of given density for which reconnection diffusion is efficient. 

If turbulence is  superAlfv\'{e}nic at the injection scale $L_{inj}$, at small scales it gets subAlfv\'{e}nic. If one approximates superAlfv\'{e}nic motions as well by Kolmogorov type scaling (given by Eq. (\ref{Kolm})), it is possible to find the constraint on $l$ for the motions to be subAlfv\'{e}nic at the scale:
\begin{equation}
l_{sub}< L_{inj} (V_A/V_{inj})^3.
\label{limit}
\end{equation}
For scales larger than this, the turbulence is superAlfv\'{e}nic and the reconnection diffusion is described by the turbulent diffusivity coefficient $\sim l v_l$, and the criterion for the 
reconnection diffusion to dominate at the scale $l$ is
\begin{equation}
l_{upper} < \left(\frac{3\pi}{32}\right)^{3/4} \frac{V_{inj}^{3/2}}{L_{inj}^{1/2}} \frac{1}{(G\rho)^{3/4}}.
\label{case2}
\end{equation}
Within the range of $[l_{lower}, l_{upper}]$, the reconnection diffusion induced by turbulence removes magnetic flux from clouds before they collapse. Beyond this range the gravitational compression of matter will increase the strength of the magnetic field
in agreement with the tendency shown in the observational data in Figure \ref{fig:zeeman}.

\begin{figure*}
\centering
 \includegraphics[height=.30\textheight]{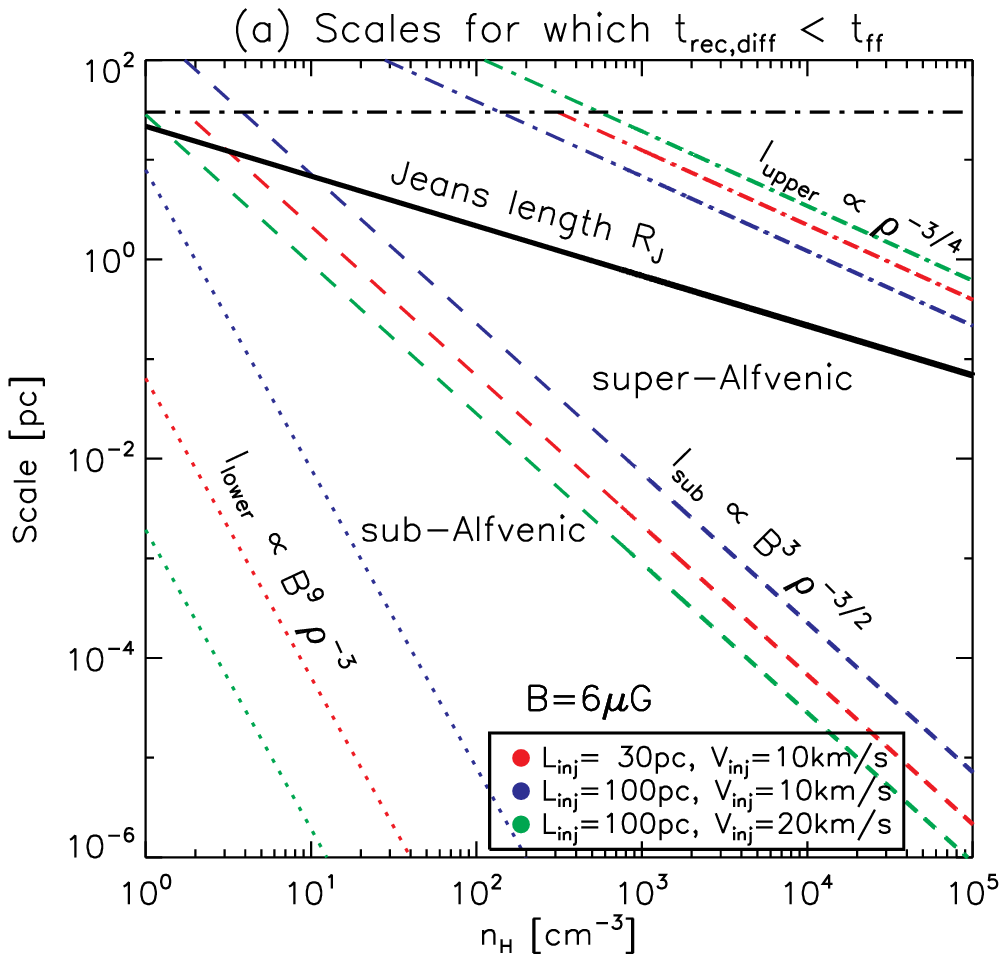}
 \includegraphics[height=.30\textheight]{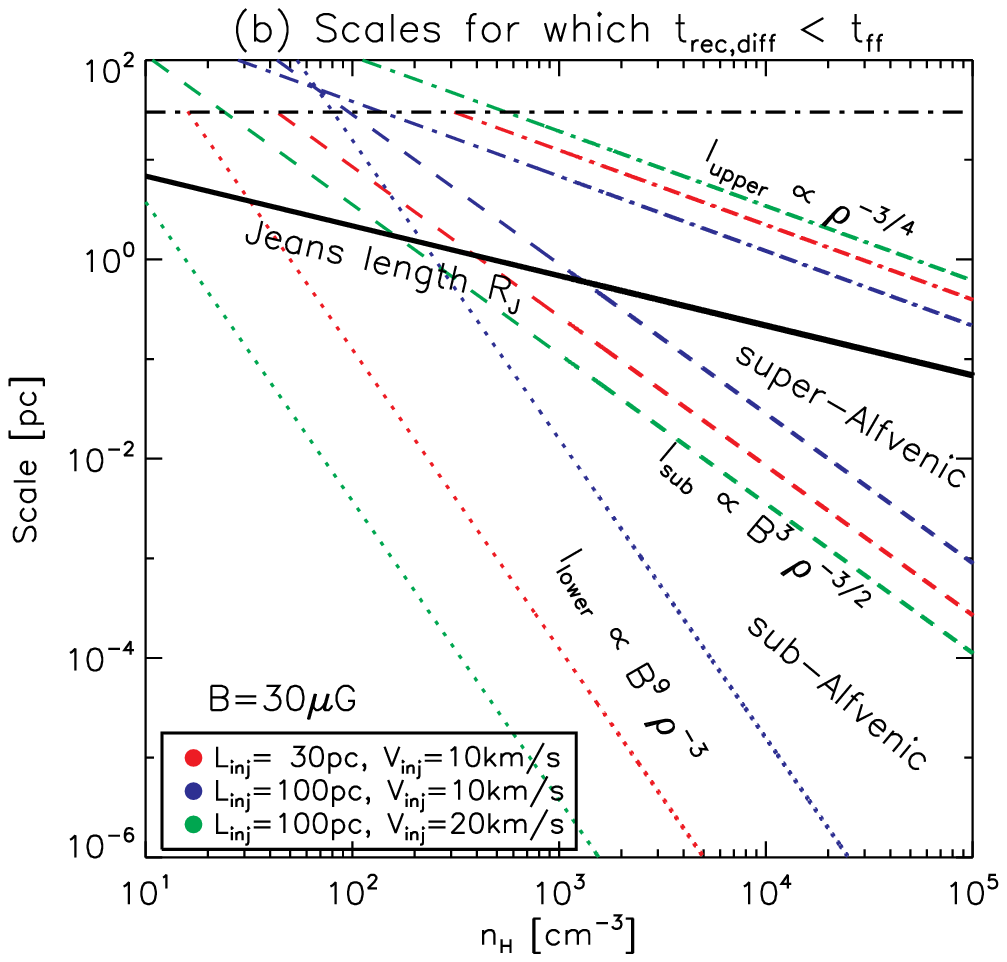}
 \caption{Illustration of the parameter space
 for which reconnection diffusion is faster than the gravitational collapse. The plots corresponds to three different models of turbulent injection described in the text. Lines corresponding to $l_{upper}$, $l_{lower}$ and the Jeans scale $R_J$ are shown. The
 horizontal lines corresponds to the injection scale of $30$ pc of Model 1.  {\bf Left panel}: Case of magnetic field of $B=6\mu\mathrm{G}$. {\bf Right panel}: Case of $B=30\mu\mathrm{G}$ magneti field. }
\label{diffvsff}
\end{figure*} 

In Figure \ref{diffvsff} we illustrate the region delimited by $[l_{lower}, l_{upper}]$ for a few selected models of turbulence injection. In particular, we consider
{\it Model} 1 with  $L_{inj}=30~\mathrm{pc}$, with $V_L=10~\mathrm{km~s^{-1}}$, {\it Model} 2 with  $L_{inj}=100~\mathrm{pc}$, with $V_L=10~\mathrm{km~s^{-1}}$ and {\it Model} 3
with $L_{inj}=30~\mathrm{pc}$, with $V_L=20~\mathrm{km~s^{-1}}$. In spite of the decades of studies, the characteristics of turbulence are not well determined. We believe that
{\it Model} 1 may be closer to reality in terms of representation of the parameters of turbulence in the galactic disk. It conservatively adopts
the scale which is smaller than the one obtained in Chepurnov et al. (2010) for atomic HI at high latitudes. In other words, it takes into account that the
scale of turbulent eddies may be smaller in the disk compared to high latitudes. In two panels of Figure \ref{diffvsff} the results are illustrated for two values of magnetization, 
namely, for $B=6$ $\mu\mathrm{G}$  and $B=30$ $\mu\mathrm{G}$. Naturally, for {\it Model} 1 the calculations are valid only up to the scales less than injection scale, which is shown
by horizontal lines in the upper parts of the panels\footnote{Therefore for {\it Model 1} the lines do not extend beyond the horizontal line.}. Figure \ref{diffvsff} also shows the regions of subAlfv\'{e}nic and superAlfv\'{e}nic turbulence. The boundary between the latter regions is given by Eq. (\ref{limit}).

For illustrative purposes we also show on the plots the Jeans scale
\begin{equation}
R_J\approx \left(\frac{15kT}{4\pi G\rho \mu m_H}\right)^{1/2},
\label{jeans}
\end{equation}
where $\mu m_H$, is the molecular mass of the gas, which is 2 in our case, and $T$ is the gas temperature $T$, which is 20K for our case. The corresponding line in Figure \ref{diffvsff} in most cases is above the lines of $l_{lower}$ and therefore no free fall happens for such small
lengths. That is why we focus in our discussion on $l_{upper}$ boundary.

According to  Figure \ref{fig:zeeman} the transition to the regime in which reconnection diffusion is slower than free fall happens at roughly 
$n\approx 10^{4}$ cm$^{-3}$. The critical cloud scale corresponding to the dominance of the free fall is given by $l_{upper}$ boundary. For Model 1 in Figure \ref{diffvsff} the scale  is around several parsecs. According to Eq. (\ref{case2}) this scales depends as $L_{inj}^{-1/2}$, which results in small variations within the range of possible injection scales. We see
that within the range of parameter variations illustrated by Figure \ref{diffvsff} the critical cloud scale does not vary significantly. Thus for our models the
column density of the relevant clouds is expected to be of the order of  $N_H\approx 10^{23}$ cm$^{-2}$. While this should be treated only as a very rough estimate due to the nature of our simplified treatment, it is encouraging that this column density corresponds to the observational data in Figure \ref{N}. 
 
In the scenario that we advocate here the clouds that undergo further collapse are supercritical and therefore magnetic fields cannot stop the
free fall. The intensity of the magnetic field is increasing, however. In reality, the approximation of the cloud collapse with $t_{ff}$ is an idealization. 
Clouds being steered by turbulence get virialized and free-falling clouds do not exist. Therefore, in agreement with Figure \ref{N}, we expect to see
dense clouds which existed long enough for reconnection diffusion to remove the excess of compressed magnetic flux. As the density increases,
gravity becomes more important and the tendency of increasing magnetic field with density gets more prominent in Figure \ref{N}.  

Our analysis above is of illustrative nature and we accept the existing uncertainties related to the injection scale and velocity as well as influence
of the compressible motions and the effect of compressions on turbulence (see Robertson \& Goldreich 2012).   Nevertheless, our results indicate the
existence of the domain $[l_{lower}, l_{upper}]$ within which the reconnection diffusion should be faster than the free fall. Outside the domain, cloud
collapse can happen more rapidly than reconnection diffusion, 
dragging the magnetic field with it. This may be responsible for the 2/3rds power-law dependence of field strength on density that we observe (Figure \ref{fig:zeeman}) for densities larger than $n \sim 10^4$ cm$^{-3}$. 

\section{Further implications of reconnection diffusion}

\subsection{Larson laws} 

We note that the observational results reviewed by Crutcher (2012) on the approximate independence of magnetic field strength on the cloud column
density up to $N_H \sim 10^{23}$ cm$^{-2}$ (and the reconnection diffusion concept that accounts for these results) can be related to the empirical Larson relations (Larson 1981) obtained for interstellar turbulence. Larson (1981) found that the velocity dispersion is proportional to the square root of the cloud size, i.e. 
\begin{equation}
\sigma_{V}\sim R^{1/2}~~,
\label{Lars1}
\end{equation}
and that the 3D density of a cloud is inversely proportional to cloud size, i.e. 
\begin{equation}
\rho \sim R^{-1}~~.
\label{Lars2}
\end{equation}
For instance, one can assume a rough equality between the kinetic energy and magnetic energy
\begin{equation}
\frac{B^2}{8\pi}\sim \rho \sigma_{v}^2~~,
\label{equality}
\end{equation}
which is true for transAlfv\'{e}nic turbulence where motions
are associated with non-linear Alfv\'{e}n modes (Goldreich \& Sridhar 1995)\footnote{The proportionality of
the two quantities is valid for superAlfv\'{e}nic  turbulence as well (see Beresnyak 2012).}.
It is also natural to accept
 the virialization of a cloud
 \begin{equation}
\frac{GM}{R}\sim \sigma_v^2~~~.
\label{virial}
\end{equation}
The Eqs. (\ref{equality}) (\ref{virial}), if combined with the simplest estimate of cloud mass  $M\sim \rho R^3$, yield
\begin{equation}
\sigma_v\sim B^{1/2} R^{1/2} 
\end{equation}
and
\begin{equation}
\rho\sim B R^{-1}.
\end{equation}
These reproduce the Larson (1981) relations given by Eqs. (\ref{Lars1}) and (\ref{Lars2}) if, in agreement with observations\footnote{We may notice the dependences of the scales on densities obtained in \S 5.2 are 
$l_{lower}>C_1/\rho^{-3}$ and $l_{upper}<C_2/\rho^{-3/4}$, where $C_1$ and $C_2$ are coefficients. The Larson law given by Eq. (\ref{Lars2}) corresponds to $l\sim \rho^{-1}$, and it fits within boundaries  determined by the upper and lower constraints on the domain of efficient reconnection diffusion.} (see Figures \ref{fig:zeeman} and \ref{N}), the reconnection diffusion keeps magnetic field uncorrelated with density. 

For cloud cores where reconnection diffusion is not fast enough to remove magnetic field on the time scale of the dynamic collapse (see
Tafalla et al. 1998, Reiter et al. 2011),
the Larson relations fail, in agreement with observations and simulations (see Nakamura \& Li 2011). Quantitatively, the
criterion for the reconnection diffusion to be able to remove the magnetic field from the collapsing cloud 
of size $l_{cloud}$ is $V_{infall}<\kappa/l_{cloud}$, where
$\kappa$ is one of the diffusion coefficients (see Eqs. (\ref{kappa_st}) or (\ref{hydro})). This is equivalent to our $t_{rec, diff}>t_{ff}$ that we discussed
in \S 5.3.

Note that the above suggestion to explain Larson (1981) laws is not unique. For instance, there are ways to explain Eq. (\ref{Lars1}) appealing to 
the properties of MHD turbulence. Indeed, for the GS95 turbulence, the turbulent velocity $v$ scales with the fourth root of the parallel size of the eddies, i.e. $v\sim l_{\|}^{1/4}$ and if, as we discussed in \S 2, the parallel scale of turbulent motions is associated with the cloud size, Eq. (\ref{Lars1})
follows trivially. Another way to reproduce the same law (i.e. Eq. (\ref{Lars1})) is to assume steep supersonic
turbulence with $E\sim k^{-2}$ in clouds, which also would result in a similar scaling\footnote{Steep
spectra of turbulence correspond to the observations of cold HI in Chepurnov et al. (2010).}. However, using the latter two approaches it
is not easy to reproduce the second relation given by Eq. \ref{Lars2}, which decreases the value of such derivations. An interesting suggestion was also made recently by Kritsuk \& Norman (2012). 
One cannot exclude that, in reality,
different mechanisms may act together providing the Larson (1981) scalings for a variety of clouds. 

At the same time, the results in Robertson \& Goldreich (2012) indicate that in order to explain the Larson (1981) laws appealing to clouds collapse, does require that the infall velocities be comparable with turbulent ones. Observations (see Tafalla et al. 1998) indicate that the cloud collapse is usually slower that that.

\subsection{Expectations for the reconnection diffusion process}

The calculations presented in Santos-Lima et al. (2010) testify that reconnection diffusion can naturally explain the poor correlation of magnetic field and density reported for the diffuse media in Troland \& Heiles (1986). This
result is rather obvious from the physical picture of reconnection diffusion that we presented above. Indeed, eddies
with different magnetization depicted in Figure \ref{mix} are expected to exchange partially ionized gas irrespectively
of the degree of gas ionization. Therefore the high ionization of media that was dealt with by Troland \& Heiles (1986) 
should not present a problem for decorrelating densities and magnetic fields. Figure \ref{fig:zeeman} testifies that
the decorrelation extends to higher densities and, moreover, H~I and lower density molecular clouds tend to have the same magnetization (equal to that in the large-scale interstellar medium) that is independent of their density. This corresponds to the expectation of the reconnection diffusion process\footnote{An
alternative explanation of the Troland \& Heiles (1986) results was provided in Passot \& Vazquez-Semadeni (2003) and it is based on the notion that MHD turbulence is a superposition of turbulent cascades of Alfv\'{e}n, slow and fast modes
(Cho \& Lazarian 2002, 2003) and therefore the enhancements of magnetic field and densities may correlate or anti-correlate within different modes. Unlike reconnection diffusion, this cannot explain the results in Figure \ref{fig:zeeman}.}. 

In Lazarian \& Vishniac (2009) it was suggested that reconnection diffusion can remove magnetic fields from protostellar accretion disks to explain observations discussed in Shu et al. (2005). Detailed MHD calculations in Santos-Lima et al. (2012) showed that the process can resolve so called ``magnetic breaking catastrophe" of the disks. Another problem where reconnection diffusion may be valuable is the so called  ``magnetic flux problem'' (see Galli et al. 2006 and references therein). For example, T-Tauri stars have magnetic fields $\approx 2\times 10^3$ Gauss (see Johns-Krull 2007) or $\Phi/M\approx 3\times 10^{-8}$ gauss~cm$^2$~g$^{-1}$, which is a million times smaller that the flux to mass ratio estimated for a one solar mass clump in a cloud of density $10^4$ cm$^{-3}$ (see Draine 2011). While further detailed quantitative modeling of the problem would be valuable, the numerical results obtained with reconnection diffusion are encouraging. For instance, the removal of magnetic flux via reconnection diffusion was reported both at the stage of turbulent cloud formation (Santos-Lima et al. 2010, Leao et al. 2012) and at the stage of the accretion disk evolution in
the presence of turbulence (Santos-Lima et al. 2012).  

It is encouraging that the reconnection diffusion concept provides new ways of approaching challenging physics of star formation in extreme environments.  For instance, galaxies emitting more than $10^{12}$ solar luminosities in the far-infrared are called ultra-luminous infrared galaxies or ULIRGs. The physical conditions in such galaxies are extreme with a very high density of 
cosmic rays (see Papadopoulos et al. 2011). Ambipolar diffusion is expected to be suppressed due to cosmic ray
ionization. At the same time these environments have the highest star formation rates, which is suggestive of a process that removes magnetic fields independent of the level of ionization. Reconnection diffusion is such a process.  

There are more cases when the predictions based on reconnection diffusion provide solutions to long standing problems. For instance, the observed star formation rate is about the same in galaxies with low metallicities as in galaxies with high metallicities (Elmegreen \& Scalo 2004). This is hard to understand if magnetic field loss is governed by ambipolar diffusion, which is supposed to be much faster in low-$Z$ galaxies with high metallicities. As a result,  contrary to observations, ambipolar diffusion theory predicts that the Initial Mass Function (IMF) and therefore the star formation rate would be shifted in low-$Z$ galaxies. As we discussed earlier, reconnection diffusion is independent of metallicity, and thus provides a solution consistent with observations. 
 
In addition, an important property of reconnection diffusion is that, unlike ambipolar diffusion, it not only provides the removal of the magnetic field, but also induces its turbulent mixing, which tends to make the distribution of magnetic field uniform. Turbulent mixing helps in keeping the diffuse media in a magnetized subcritical state. In this situation, the external pressure is important for initiating collapse,  which well corresponds to the observations of numerous small dark clouds not forming stars in the inter-arm regions of galaxies (Elmegreen 2011).    

Finally, the reconnection diffusion concept implies that in realistic turbulent media there is no characteristic density for the collapse to be initiated. Therefore any cloud with the appropriate virial parameter (see McKee \& Zweibel 1992) can form stars. The difference between different clouds stems from the density controlling the
timescale of the collapse and the level of steering provided by turbulence or temperature of the media. Directly, the requirement of clouds to be molecular is not present for the reconnection diffusion to induce star formation. However, molecular clouds have lower temperature. 

\subsection{Mean magnetic field and reconnection diffusion}

The reconnection diffusion process, as any diffusion process, is expected to relax the sharp local changes of magnetic field direction. However, as reconnection diffusion is associated with turbulence, the latter introduces variations of magnetic field direction. 

In terms of reconnection diffusion some observations are definitive, some are suggestive. For instance, we believe that reconnection diffusion can explain the alignment of magnetic field of molecular clouds cores with the magnetic field of the spiral arms (see the example of M33 measurements in Li \& Henning 2011). Indeed, it seems natural that if molecular clouds are self-gravitating, the gravitational contraction associated with the cloud formation is expected to distort the local direction of magnetic field even in the case of subAlfv\'{e}nic turbulence
of the ordered galactic magnetic fields (see Eq. (\ref{LV99})). Similarly, the distortions are expected to arise from the molecular clouds sharing its 
angular momentum with the ambient media during their contraction.
Whatever the corresponding Alfv\'{e}nic Mach
number $M_A$ the reconnection diffusion can smooth and relax the structure of the otherwise distorted mean magnetic field. We have the extensive evidence of the molecular clouds being turbulent (see McKee \& Ostriker 2007 and ref. therein) and therefore the process of reconnection diffusion there is inevitable. This process
will tend to move magnetic in molecular clouds correlated with the mean magnetic field removing the distortions induced by the formation process\footnote{In view of the discussion above  an additional explanation is required. Why would not reconnection substantially distort
the original direction of the magnetic field? Indeed the visualization of simulations of the reconnection 
in the presence of turbulence (see Lazarian, Eyink \& Vishniac 2012) show substantial changes of the magnetic
field direction introduced by magnetic reconnection. However, these changes are in the component where magnetic field reverses its direction. In the case of reconnection diffusion we are dealing with the magnetic fields that may be originally at a small angle to each other (see Figures \ref{recdiff} \ref{mix}). It is evident from topological considerations that reconnection of such fields cannot create a substantial change of the observed magnetic field direction (see also the discussion in \S2).}. An alternative interpretation of the results of Li \& Henning (2011) is that the interstellar matter is collected
along magnetic field lines weakly distorted by subAlfv\'{e}nic turbulence. While the quantitative analysis of the latter suggestion is beyond the scope of the present work, we would like to stress that the magnetic reconnection is arguably an essential part of subAlfv\'{e}nic magnetic turbulence (see LV99, Eyink et al. 2011). Therefore numerical simulations of the accumulation of matter along magnetic field lines within subAlfv\'{e}nic turbulence do intrinsically include the magnetic field smoothing by reconnection diffusion. While we accept that without a detailed study the observations by Li \& Henning (2011)
do not provide a conclusive test of reconnection diffusion, we feel that reconnection
diffusion provides a viable explanation consistent with the observations. 

Another possible explanation of molecular clouds having magnetic field correlated with the mean galactic magnetic field is related to the accumulation of matter along magnetic field lines. With our paper being focused on the comparison of the predictions of reconnection versus ambipolar diffusion we leave a detailed comparison of the reconnection diffusion with the aforementioned non-diffusive scenario of matter collection for further publications. We may mention parenthetically that reconnection diffusion is expected to affect significantly such a collection (see Appendix D).

\section{Discussion}
\subsection{Reconnection diffusion: simulations and estimates}

This paper was motivated by the observational results reviewed by Crutcher (2012), who showed that several important characteristics were not compatible with the ambipolar diffusion theory. We have shown that reconnection diffusion can produce agreement with the observations. Reconnection diffusion is
based on the numerically tested LV99 model of fast reconnection in turbulent media as well as more recent insights into the violation of the frozen-in condition in turbulent magnetized fluids (Eyink 2011, ELV11). 
It is also appeals to the direct 3D MHD simulations of the magnetic field diffusion in Santos-Lima et al. (2010, 2012) that showed consistency with the statistical description of the magnetic diffusion in turbulent fluids (see Lazarian 2006). Therefore in the present paper we use this statistical description
directly. This allows us to get insight into the evolution of magnetic fields at the core and the envelope.

The approach to studying reconnection diffusion that is based on 3D numerical simulation has its own limitations. In Santos-Lima et al. (2010) we used an external gravitational potential. In more recent simulations in Leao et al. (2012) self-gravity is used. However, using an isothermal code to make the turbulent cloud more well-defined in the absence of the confining pressure of the ambient gas at a different temperature, we imposed an additional gravitational potential. We are experimenting with the
ideas of introducing the ambient pressure without compromising the resolution and the simplicity of
the simulation interpretation, but this is work in progress. In this situation, it is advantageous to 
get estimates and analyze idealized situations with clear physical meaning the way we do in this paper. Naturally, our simplified calculations do not seek the quantitative agreement with the observations, but only illustrate the qualitative
behavior. Indeed, within the adopted toy model we do not evolve density together with magnetic field, but consider the diffusion of magnetic
field out of an idealized cloud of constant density\footnote{A possible initial set up with the cloud of gas density $\sim 60$ cm$^{-3}$, radius 2.5 pc and the initial magnetic field of $\sim 6\times 10^{-6}$G may be advantageous. Such a cloud would be subcritical, which agrees with the notion of diffuse H~I clouds being subcritical. Then the reconnection diffusion would remove flux such that the  cloud could collapse, with density building up to $10^4$ cm$^{-3}$ in the $r=0.2$ pc core and up to $10^3$ cm$^{-3}$ in the $r=1$ pc envelope, which are the mean parameters of the clouds observed by Crutcher, Hakobian \& Troland (2009). This is what we will
try in future.}.   

We can mention that it is possible to argue that the process of reconnection diffusion is present in some of high resolution numerical simulations. However, without clear identification of the role of turbulence in fast reconnection, one may not be sure when the results are due to physically motivated reconnection diffusion and when they are
the consequence of the bogus effects of numerical diffusion. For instance, Crutcher, Hakobian \& Troland (2009) refer to the simulations in Luntilla et al. (2009) that produce, in 
agreement with observations, higher magnetization of the cloud cores. If these cores are of the size of several grid units across, numerical
effects rather than reconnection diffusion may be dominant and turbulence is suppressed at these scales.

Our results show that in the presence of reconnection diffusion it is natural to expect that the magnetic
field diffuses faster from the area of the envelope and the magnetic field strength is larger at the cloud 
core. This conclusion does not really depend on the particular model of turbulence and its distribution
in the molecular cloud as far as those are constrained by existing observations. 

We should also stress that the 3D numerical simulations like those in Santos-Lima et al. (2010, 2012) {\it by themselves} do not provide the description
of reconnection diffusion in realistic astrophysical environments. The interpretation of the results requires the proper understanding of scaling of 
magnetic reconnection with the dimensionless combination called the Lunquist number $S\equiv (L_{cur. sh} V_A/\eta)$, where $L_{cur. sh.}$ is the
extent of the relevant current sheet, $\eta$ is Ohmic diffusivity. The Lundquist numbers in molecular clouds and in the corresponding simulations
differ by a factor larger than $10^5$. In this situation one can establish the correspondence between the numerical simulations and astrophysical reality
only if the reconnection does not depend on $S$. The independence from $S$ of magnetic reconnection is the conclusion of LV99 model. This model
has been tested in Kowal et al. (2009, 2012) via a set of dedicated numerical simulations that confirmed the scaling predictions in LV99. This work
exemplifies the advantages of the synergy of scaling arguments with numerical simulations as opposed to the ``brute force numerical approach'', which may not be productive while dealing with turbulence.

\subsection{Predictions of the reconnection and ambipolar diffusion models}

Within the concept of ambipolar diffusion the explanation of observational results reviewed by Crutcher (2012) is extremely difficult. Our point is that
ambipolar diffusion is not the only process that can be responsible for the removal of magnetic flux from the media. We show that reconnection diffusion can reduce the magnetic field strength in the envelope with respect to the core. On a more
fundamental level, reconnection diffusion changes the mass to flux ratio allowing magnetic field to diffuse away from the center of the gravitational potential. 

The available observations provide a rather mild constraint, namely, that mass/flux does not increase as fast in the core as ambipolar diffusion predicts. In many ways, ambipolar diffusion is a very special type of diffusion, the efficiency of which drops towards more ionized outer regions of the envelope inducing more efficient flux loss from cores.  

Typically the observed column densities through the cores studied by Crutcher et al. (2009) are about twice those through the envelope regions. If mass/flux were constant, then the fields in the envelopes would be 1/2 those in the cores. So just saying that the fields in the envelopes are less than those in the cores does not require reconnection diffusion, or any kind of diffusion. Reconnection diffusion is required if the envelope fields are even weaker than half of the core fields; this is what is observed. Ambipolar diffusion would require that envelope fields are stronger than half of the core fields. 

\subsection{Reconnection diffusion and ambipolar drift in turbulent media}

Within this study we do not appeal to ambipolar diffusion, which is acceptable when the reconnection
diffusion provides larger diffusivity for magnetic fields. One may argue that this is a generic
situation in the presence of turbulence.

For instance, Heitsch et al. (2004, henceforth HX04) performed 2.5D simulations of turbulence with two-fluid code and examined the decorrelation of neutrals and magnetic field
that was taking place as they were driving the turbulence. The study reported an enhancement of the ambipolar diffusion rate compared
to the ambipolar diffusion acting in a laminar fluid. HX04 correctly associated the enhancement with turbulence
creating density gradients that are being dissolved by ambipolar diffusion (see also Zweibel 2002).
While in 2.5D simulations of HX04 the numerical set-up precluded reconnection from happening (as magnetic
field was perpendicular to the mixing plane and magnetic fields never crossed each other at an angle), the
authors reported an enhanced rate that is equal to the turbulent diffusion rate $L V_L$, which is the
result expected as a limiting case of the reconnection diffusion prediction for the special set-up studied.

While we agree with HX04 in terms of the importance of turbulence, we suggest that it is misleading to refer to the process as ``turbulent ambipolar
diffusion''. We believe that ambipolar diffusion does not play a role in the enhancement measured and the
observed effect is entirely due to turbulence\footnote{A similar process takes place in the case of molecular diffusivity in turbulent hydrodynamic flows. The result for the latter flows is well known: in the turbulent regime, molecular diffusivity is irrelevant
for the turbulent transport. Indeed, in the case of high microscopic diffusivity, turbulence provides mixing down to a scale $l_1$ at which the microscopic diffusivity both suppresses the cascade and ensures efficient diffusivity of the contaminant. In the case of low microscopic diffusivity, turbulent mixing happens down to a scale $l_2\ll l_1$, which ensures that even low microscopic diffusivity is sufficient to provide efficient diffusion. In both cases the total effective diffusivity of the contaminant is given by the product of the turbulent injection scale and the turbulent velocity.}. We note that, in the presence of turbulence, the independence of the gravitational collapse from the ambipolar diffusion rate was reported in numerical simulations by Balsara, Crutcher \& Pouquet (2001). 

Ambipolar diffusion may control flux removal from the laminar cores. In the presence of turbulence, it is reconnection diffusion that dominates. The
results of HX04 support this notion, as the authors in their set up report the transition to turbulent diffusivity in the presence of turbulence. The set-up in
HX04 precludes the magnetic fields from reconnection as field lines perform mixing motions being kept absolutely parallel. In any realistic turbulent 3D flow magnetic reconnection will be essential.   

\subsection{Reconnection diffusion and hyper-restivity concept}

Another process of diffusive nature is related to ``hyperresistivity'' or enhanced physical resistivity of turbulent fluids. To explain fast removal of magnetic field from accretion disks Shu et al. (2006) appealed to the hyperrestivity concept (Strauss 1986, 1988, Bhattacharjee \& Hameiri 1986, Hameiri \& Bhattacharjee, Diamond \& Malkov 2003). The studies introducing hyper-resistivity attempt to derive the effective resistivity of the turbulent media in the context of the mean-field resistive MHD. Using magnetic helicity conservation the authors derived the electric field. Then, integrating by part, they obtained a term which could be identified with effective resistivity proportional to the magnetic helicity current. There are several problems with this derivation. In particular,
the most serious is the assumption that the helicity of magnetic field and the small scale turbulent fields are separately conserved, which erroneously disregard the magnetic helicity fluxes through open boundaries that is essential for fast stationary reconnection (see ELV11). 
In more general terms, hyper-resistivity idea is an incarnation of the mean-field approach to producing fast reconnection. As explained in ELV11, the problem of such approaches is that the lines of the actual astrophysical magnetic field should reconnect, not the lines of the mean field. Therefore the correct approach to fast reconnection should be independent of spatial and
time averaging.
   
All in all, we believe that the concept of hyper-resistivity should not be applied to astrophysical environments. 

\subsection{Reconnection diffusion and ``magnetic turbulent diffusivity'' concept}

One may claim that the reconnection diffusion concept extends the concept of hydrodynamic turbulent diffusion to magnetized fluids. The physics of it is very different from the ``magnetic turbulent
diffusivity'' idea discussed within the theories of the kinematic dynamo (see Parker 1979). Reconnection diffusion, unlike ``magnetic turbulent diffusivity'', deals with dynamically important
magnetic fields, e.g. with subAlfv\'{e}nic turbulence\footnote{The concept of ``magnetic turbulent diffusivity'' assumes that magnetic fields can be passively
mixed up on all scales, up to the Ohmic diffusion scale. At the latter scale Ohmic dissipation is trivial and the diffusion is hydrodynamic at all scales.}. Thus in the process of reconnection diffusion, magnetic fields are {\it not} passively mixed and magnetic reconnection plays a vital role for the process.

One may show that the domain of applicability of the ``magnetic turbulent diffusivity'' concept is extremely limited. Numerical calculations (Cho et al. 2010,
Beresnyak 2012) testify that around 5\% of turbulent energy flow is transferred to magnetic energy and that this value does not depend on the initial magnetization of the fluid. Therefore, even if the turbulent flow initially is not magnetized, it is expected to develop dynamically important magnetization on the time scale of 10-20 large eddy turnover times. As a consequence, we expect that star formation will happen in magnetized media even for the first
generation of stars.\footnote{Incidentally, the same logic suggests that the proposals of bringing magnetic field
and kinetic energy to a state of equipartition in molecular clouds during their formation (see Sur et al. 2012) may not work. Indeed, molecular clouds are not likely to survive 20 crossing times.}

\subsection{Reconnection diffusion and numerical effects}

One might argue that reconnection diffusion is automatically a part and parcel of 3D numerical simulations of star formation. As we argue
below this is only partially true and a proper understanding of reconnection diffusion is required for the interpretation the numerical simulations.

The LV99 model predicts that the reconnection rates in turbulent fluids are independent of the local physics, but are determined by the turbulent motions. This is great news for numerical simulations: parasitic numerical effects that induce poorly controlled small scale diffusivity of magnetic field lines are not important on the scales of turbulent
motions. In other words, the low resolution numerics may provide an adequate representation of high Lundquist
number astrophysical turbulence as far as the reconnection diffusion is concerned. On the contrary, if the structures studied in numerical simulations
(e.g. cores, filaments, shells) lose turbulence due to numerical diffusivity effects, we predict that the diffusion of magnetic flux in those simulated structures differs significantly from the diffusion in the actual interstellar structures where turbulence persists. Note, that naive convergence studies based on increasing the numerical resolution several times cannot notice the problem unless the resolution increased to the degree that the aforementioned
structures become turbulent.

In terms of numerical studies of star formation it is encouraging that it is the largest scales of turbulent motions that are important for the reconnection
diffusion. At the same time, in many cases in numerical simulations we expect to observe magnetic field evolution influenced by strongly damped
turbulent motions, which do not demonstrate power law cascade. The differences of reconnection diffusion in such damped regimes and in the
astrophysically important regime of well developed turbulence with a power law spectrum should be a subject of further studies.

At the same time, the importance of understanding reconnection diffusion is surely not limited by the proper interpretation of numerical simulations of star formation. Reconnection diffusion provides a theoretical way to predict the dynamics of magnetic flux in turbulent fluids.

\subsection{Magnetic star formation in the presence of reconnection diffusion}

In spite of the fact that reconnection diffusion relaxes the conventional flux freezing condition, the role of
magnetic fields for star formation should not be disregarded. For instance, magnetic fields solve the angular
momentum problem in the process of star formation (Mestel 1965). At the same time, reconnection diffusion
allows to prevent the ``magnetic breaking catastrophe'' that, contrary to observations, prevents the formation
of circumstellar disks (Lazarian \& Vishniac 2009, Santos-Lima et al. 2012). 
 
 \subsection{Comparison with earlier papers}
 
 The concept of removal of magnetic fields from clouds via the process of LV99 reconnection was first introduced in Lazarian (2005) (see also
 discussions in Lazarian \& Vishniac 2009).  The
 numerical calculations of reconnection diffusion in application to molecular clouds and accretion disks have been performed in Santos-Lima et al. (2010, 2012). However, the physical picture of how magnetic fields and matter can decouple in turbulent magnetized plasma was not presented there. 
 
 In comparison, this paper focuses on the physics of reconnection diffusion and presents a simple analytical model illustrating how the tendencies 
 in magnetic field distribution revealed in the observations by Crutcher et al. (2009, 2010a) can be accounted for, as well as other observations reviewed by Crutcher (2012). In addition, the current paper presents the additional evidence of the importance of reconnection diffusion in various astrophysical environments. While alternative explanations may be possible for the cases discussed (e.g, Li et al. 2011) we feel that the consistency of those with the reconnection diffusion concept provides an additional evidence in favor of the reconnection diffusion scenario. This calls for the necessity to include
 the concept of reconnection diffusion within the star formation paradigm.

\section{Summary}

This paper presents the physical justifications of the concept of reconnection diffusion and an analytical toy model as 
well as simple numerical calculations of the effects of the process on the magnetic flux loss out of clouds. 
The major results of this paper can be briefly summarized in the following way:
\begin{itemize}
\item Reconnection in turbulent media happens fast, and this enables turbulent diffusion of matter
and magnetic field. This process is termed reconnection diffusion, and it provides a way to
change the flux to mass ratio of molecular clouds.
\item The process of reconnection diffusion does not depend on the ionization of the cloud material;
therefore, if the magnetic field is captured within a collapsing turbulent cloud, the decrease of the 
magnetic field strength due to reconnection diffusion is most prominent at the edges of the cloud.
This is a robust result that does not depend on the details of the turbulence picture.
\item The predictions of reconnection diffusion are consistent with the measurements of the
ratio of the magnetization of cores and envelops of molecular clouds obtained in Crutcher et al. (2009), 
with the poor density -- magnetic field correlation at low densities, with the 2/3rds power law correlation at high densities reported in Crutcher et al. (2010b), and with the large range in magnetic field strengths in molecular clouds of the same density. 
\item In agreement with observations, reconnection diffusion predicts the value of critical column density at which the increase of magnetic field with density is measured.
Reconnection diffusion may also provide an
explanation of the well-known empirical Larson relations. 
\item Reconnection diffusion can provide a way of approaching other problems where ambipolar diffusion is
unlikely to produce viable solutions, e.g. the problems (to name a few) of poor correlation of density and magnetic field in 
diffuse media, efficient star formation in ultra-luminous infrared galaxies, alignment of magnetic field in
cloud cores with the magnetic field in spiral arms.
\item We do not argue here that reconnection diffusion is the only important MHD process that may affect star formation. If subcritical, self-gravitating clouds are formed that have weak turbulence, ambipolar diffusion would operate over longer time scales to drive the cores to supercritical states and allow stars to form. It is also possible that collection of matter along magnetic flux tubes play a role, perhaps especially at smaller spatial scales where GMCs fragment to form individual self-gravitating clouds. It seems likely that turbulence would prevent the requisite laminar flow, and reconnection diffusion would come into play. 
\end{itemize}

\appendix

\section{A: Derivation of LV99 reconnection rate}

A picture of two flux tubes of different directions which get into contact in 3D space is a generic framework to
describe magnetic reconnection. The upper panel of Figure \ref{LV} illustrates why reconnection is so slow in the textbook Sweet-Parker model. Indeed, one can write
\begin{equation}  
v_{rec}= V_A \frac{\Delta}{L_x}, 
\label{vrec} 
\end{equation}
meaning that $v_{rec}\ll V_A$ if $\Delta \ll L_x$.
Consider the Ohmic diffusion of magnetic field lines (see ELV11). The mean-square 
vertical distance that a magnetic field line can diffuse by resistivity in time $t$ is 
\begin{equation}
\langle y^2(t)\rangle \sim \lambda t, 
\label{diff-dist} 
\end{equation}
where $\lambda=\eta c^2/4\pi$ is Ohmic diffusivity.
The field lines are advected out of the sides of the 
reconnection layer of length $L_x$ at a velocity of order $V_A.$ Thus, the time that the lines can 
spend in the resistive layer is the Alfv\'{e}n crossing time $t_A=L_x/V_A.$ Thus, field lines can only 
be merged that are separated by a distance 
\begin{equation}
\Delta = \sqrt{\langle y^2(t_A)\rangle} \sim \sqrt{\lambda t_A} = L_x/\sqrt{S},
\label{Delta} 
\end{equation}
where $S$ is Lundquist number,
\begin{equation}
S=L_x V_A/\lambda.
\label{Lun}
\end{equation}
Combining Eqs. (\ref{vrec}) and (\ref{Delta}) one gets the famous Sweet-Parker reconnection rate, 
\begin{equation}
v_{rec, SP}=V_A/\sqrt{S}.
\label{SP}
\end{equation} 

The LV99 model generalizes the Sweet-Parker one by accounting for the existence of magnetic field line stochasticity (Figure \ref{LV} [lower panels]). In ELV11 the LV99 reconnection rate was rederived appealing to the concept
of Richardson (1926) diffusion. Richardson diffusion (see Kupiainen et al. 2003) implies the mean squared separation of particles
\begin{equation}
\langle |x_1(t)-x_2(t)|^2 \rangle\approx \epsilon t^3,
\label{Rich}
\end{equation}
 where $t$ is time, $\epsilon$ is the energy cascading rate and $\langleÉ\rangle$ denote an ensemble averaging. For subAlfv\'{e}nic turbulence $\epsilon\approx u_L^4/(V_A L)$,
where $u_L$ is the injection velocity and $L$ is an injection scale (see LV99). Therefore analogously to Eq. (\ref{Delta}), one can write
\begin{equation}
\Delta\approx \sqrt{\epsilon t_A^3}\approx L_x(L_x/L)^{1/2}M_A^2,
\label{D2}
\end{equation}
where it is assumed that $L_x<L$. Combining Eqs. (\ref{vrec}) and (\ref{D2}),
one recovers the LV99 expression for the rate of magnetic reconnection (see also ELV11),
\begin{equation}
v_{rec, LV99}\approx V_A (L_x/L)^{1/2}M_A^2,
\label{LV99ap}
\end{equation}
in the limit of $L_x<L$. Analogous considerations allow one to recover the LV99 expression for $L_x>L$, which differs 
from Eq.~(\ref{LV99}) by the change of the power $1/2$ to $-1/2$.

\section{B: Microscopic picture of reconnection diffusion}

\begin{figure}
\centering\includegraphics[height=.22\textheight]{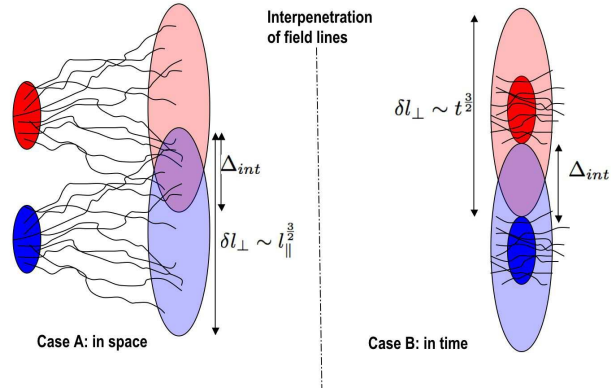}
  \caption{Microscopic physical picture of reconnection diffusion. Magnetized plasma from two regions is spread
  by turbulence and mixed up over $\Delta_{int}$. {\it Left panel}: Description of the process in terms of field wandering in space. {\it Right panel}. Description of the magnetic field line spread with time.}
 \label{alt}
\end{figure}

Magnetic field wandering provides a possibility of explaining reconnection diffusion without appealing
to magnetic eddies. According to LV99, the transverse displacement of magnetic field lines over distance $x$ is a spatial random walk given by the equation
$d\langle y^2\rangle/dx\sim L (V_L/V_A)^4$, which results in
\begin{equation}
\langle y^2 \rangle\sim L x (V_L/V_A)^4.
\label{y2}
\end{equation}
Figure \ref{alt}, left, illustrates the spread of magnetic field lines in the perpendicular direction as magnetic field lines are traced by particles
moving along them. Using Eq. (\ref{y2}) one can get the RMS separation of the magnetic field lines (LV99)
\begin{equation}
\delta l_{\bot}^2\approx \frac{l_{\|}^3}{L}\left(\frac{V_L}{V_A}\right)^4,
\label{l_rms}
\end{equation}
i.e. $l_{\bot}^2$ is proportional to $l_{\|}^3$. This regime identified in LV99 is a Richardson diffusion
but in terms of magnetic field lines\footnote{This regime induces perpendicular ``superdiffusion'' in terms of cosmic rays and other charged
particles streaming along magnetic field lines.}. The numerical testing of this prediction of LV99 is presented, e.g. in Lazarian et al. (2004). This is the case
of diffusion in space. Eq. (\ref{l_rms}) allows one to calculate the distance $l_{\|}$ at which the magnetic field lines of regions
separated by $l_{\bot}$ start overlapping\footnote{Incidentally, in terms of reconnection, Eq. (\ref{l_rms}) expresses the thickness of the outflow 
denoted as $\Delta$ in Figure \ref{LV}. Substituting this value in Eq. (\ref{vrec}), one recovers the LV99 reconnection rate, i.e. Eq. (\ref{LV99}).},
i.e. $l_{\|}\approx (\delta l_{\bot}^2 L)^{1/3}(V_A/V_L)^{4/3}$. Thus for sufficiently large $l_{\|}$ all parts of the volume of magnetized plasmas
get connected. In other words, the entire volume becomes accessible to particles moving along magnetic field lines. Naturally, in this situation the star formation community's
customary notion of flux to mass ratio loses its original sense.   

In addition, in turbulent plasmas, our turbulent volumes spread due to Richardson diffusion with $\delta l_{\perp}^2\sim t^3$. This process is
illustrated by Figure \ref{alt}, right. The magnetized plasma spreads by subAlfv\'{e}nic turbulence over a larger volume, while motions of plasma along magnetic field lines are ignored. The two initially disconnected volumes get overlapped, and the magnetic fields and plasmas of two regions get mixed up. Again, one easily can see that the process that we described in terms of magnetic field lines is similar to the one we described in terms of reconnected magnetic field flux tubes 
in Figure \ref{mix}. This is expected, as magnetic reconnection is an intrinsic part of the picture of MHD turbulence that
governs the dynamics of magnetic field lines (LV99)\footnote{One can also claim that spontaneous stochasticity of magnetic fields in turbulent fluids is an underlying process that governs magnetic reconnection (ELV11). As
we mentioned earlier, fast reconnection makes MHD turbulence theory self-consistent.}

\section{C: Tracing magnetic field lines with ions}

Star formation theory was developed in the assumption that ions trace magnetic fields. Thus two ions on the same magnetic field line are expected to stay on the same line unless their motion is perturbed by collisions. This notion needs to be modified in the presence of turbulence. 

\begin{figure*}
\centering
  \includegraphics[height=.29\textheight]{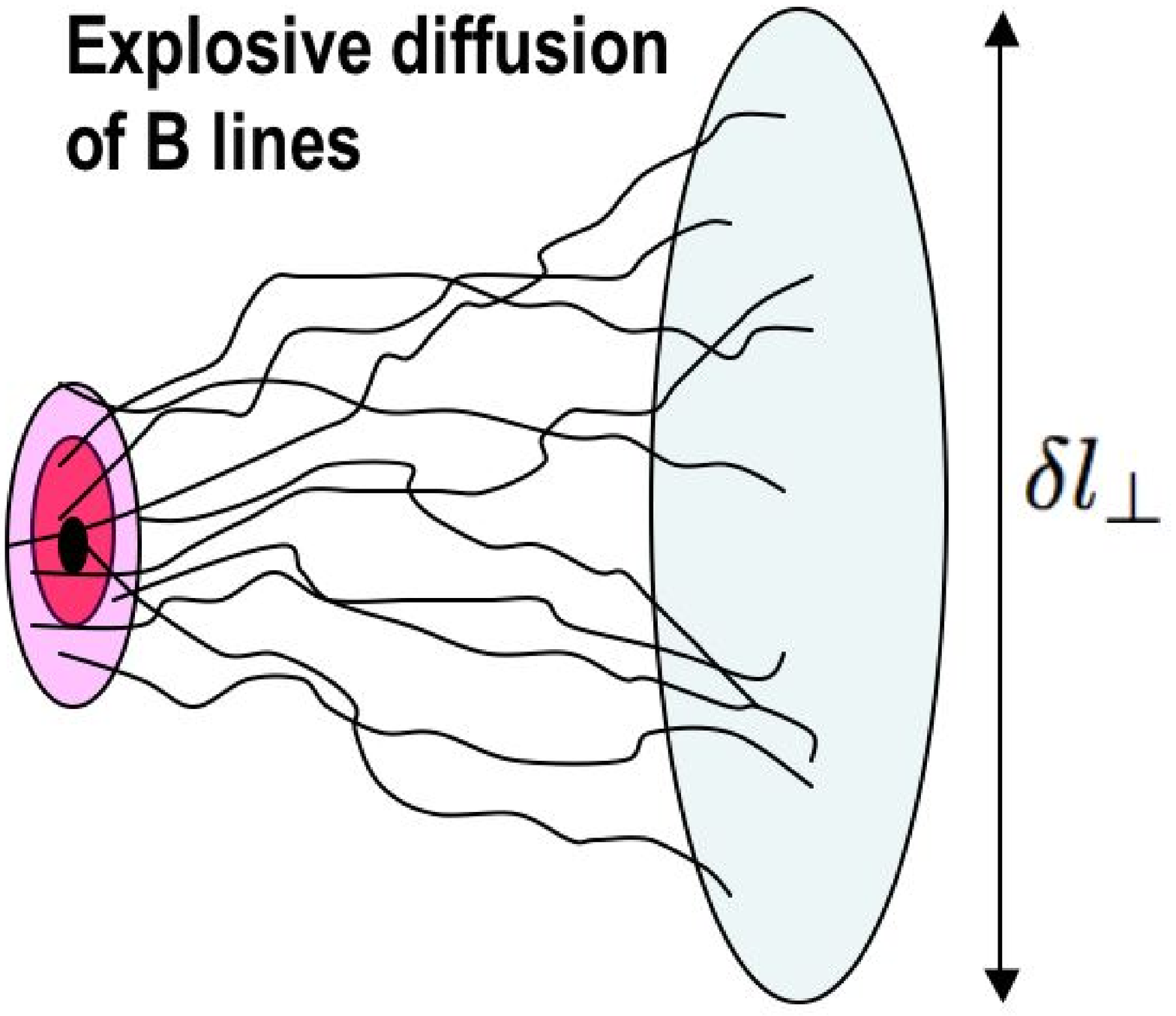}
  \includegraphics[height=.40\textheight]{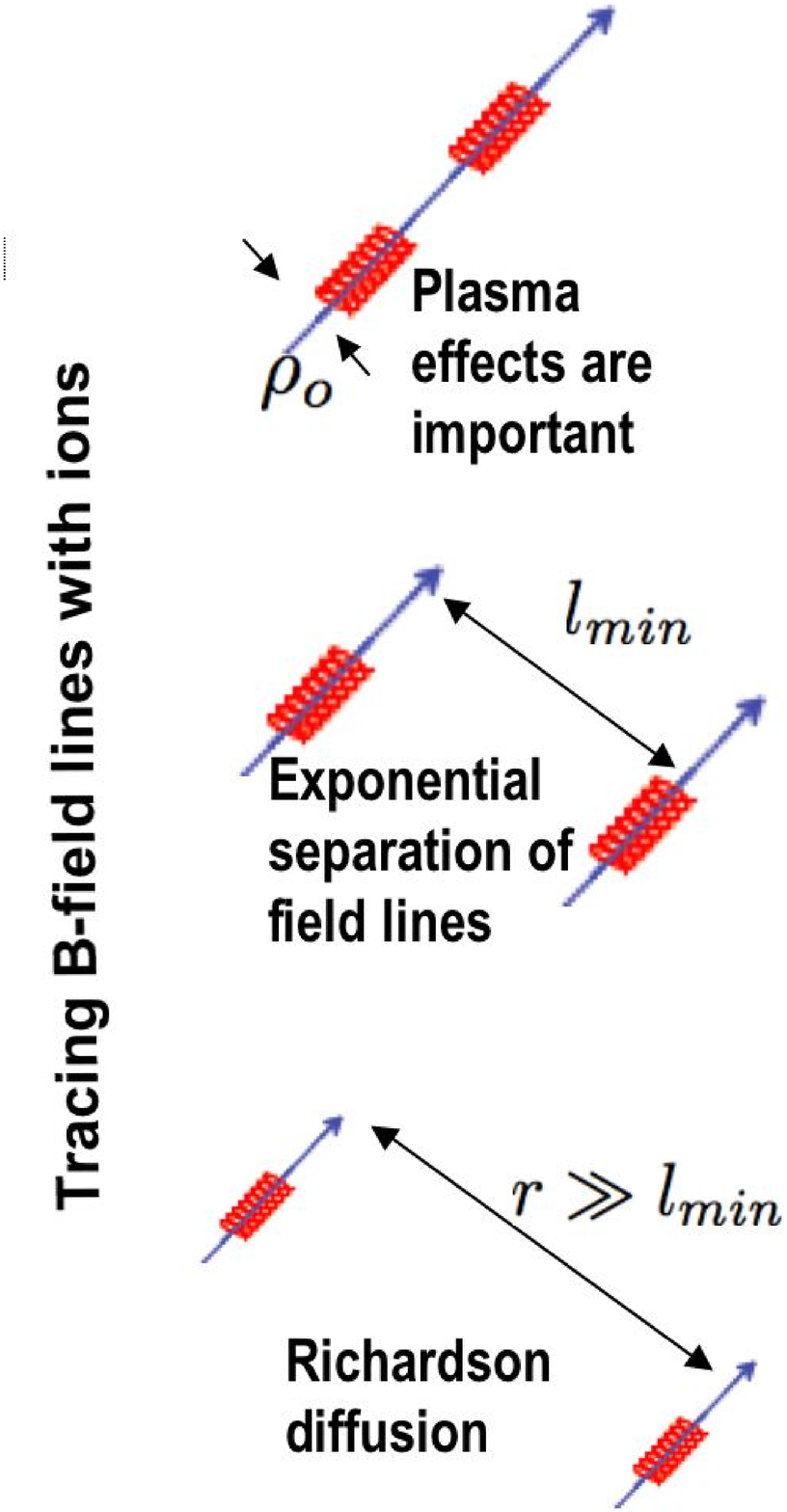}
  \caption{{\it Left Panel}: The extent of the diffusion of magnetic field lines is independent of their initial position in agreement with the concept of Richardson diffusion (see Eq. (\ref{sol})). {\it Right Panel}:
 {\bf Upper plot} Ions tracing the same magnetic field line. The diffusion and decorrelation arises
  from plasma or Ohmic effects as well Rechester-Rosenbluth effect. {\bf Middle plot}: Ions separated by scales 
  much larger than the ion Larmor radius are further separated by the Rechester-Rosenbluth effect. {\bf Lower plot}: At scales larger than the turbulence damping scale the Richardson diffusion takes over resulting in explosive
  separation of field lines.}
 \label{regimes}
\end{figure*}

The study in LV99 revealed the Richardson-type diffusion of magnetic field lines. Figure \ref{regimes} illustrates the loss of the Laplacian determinism for magnetic field lines (see also Eq. \ref{sol}). In analogy with the illustrative example above, the final line spread $l_{\bot}$ does not depend on the initial separation of the field lines. This is a remarkable effect that provides a microscopic picture of reconnection diffusion based on the description of magnetic field lines rather than on the reconnection of well-organized flux tubes.

We shall consider tracing magnetic field lines in the realistic turbulence situation with the dissipation scale $l_{min, \bot}$,
 where, as everywhere in the paper, $\bot$ denotes the scale perpendicular to the local magnetic field (see
Figure \ref{regimes}). While a detailed discussion of the Richardson type diffusion in the presence of magnetic
field is given in ELV11, we feel that our introduction of the finite dissipation scale $l_{min, \bot}$
 helps to avoid some of the paradoxes discussed in ELV11. It is also important that for the problems of
star formation $l_{min, \bot}$ may be much larger than ion Larmor radius. In other ways we follow the logic
presented in ELV11.
 First of all, resistivity, whatever its nature, introduces stochastic forcing in the description of magnetic field line dynamics. Indeed, the induction equation with the resistive term $\eta\Delta B$ signifies stochasticity associated with Ohmic diffusion. Therefore the definition of the magnetic field line on scales affected by resistivity is not deterministic. In addition, it was stressed in ELV11 that the magnetic field line motion is a concept defined by convention and not testable experimentally. This point was discussed in the literature (see Newcomb 1958,
 Vasylianas 1972, Alfv\'{e}n 1976), but sometimes forgotten. Thus magnetic field lines may be tagged by ions that start at the same field line (see Figure \ref{regimes}). In the case of smooth laminar magnetic field and ideal MHD equations the motions of ions will reveal magnetic field lines and two ions on the same field line will always remain on the same line. The situation is radically different in the presence of turbulence.

In our thought experiment we shall trace ions moving with the same velocity and
separated perpendicular to magnetic field by a distance of a Larmor radius $\rho_0$. As we discuss further, if the separation is
less $\rho_0$ than that one can appeal to plasma effects to increase the separation to $\rho_0$. Let us assume that 
the minimal scale of turbulence $l_{min, \bot} >\rho_0$. In this situation the dynamics of ions
 can be approximated by the dynamics of charged particles in toy model of ``a single scale MHD turbulence'' discussed in Rechester \& Rosenbluth 
 (1978). Indeed, the turbulent motions at the critically damped scale $l_{min, \bot}$ are dominant for shearing and steering matter and magnetic field on the smaller scales. The Rechester \& Rosenbluth (1978) theory predicts the Lypunov growth 
of the perpendicular separation of
 ions, i.e. the separation gets $\rho_0\exp(l/l_{min, \|}$, where $l$ is the distance traveled by ions and $l_{min, \|}$
 is the scale parallel scale of the critically damped eddies with the perpendicular scale $l_{min, \bot}$ (see
 also Narayan \& Medvedev 2001, Lazarian 2006). Thus to get separated by the distance $l_{min, \bot}$ the ions 
 should travel the so-called Rechester-Rosenbluth distance
 \begin{equation}
 L_{RR}\approx l_{min, \|} ln(l_{min, \bot}/\rho_0),
 \label{RR}
 \end{equation}
 which is at most a dozen times larger than the microscopic scale $l_{min, \|}$. As soon as ions get separated over
 the distance of $l_{min, \bot}$, they get into different eddies and the process of Richardson diffusion starts. 
 The separation of ions increase fast with their initial position quickly forgotten (see the solution given by
Eq. (\ref{sol}) and the detailed discussion in ELV11). Thus
 after a relatively short period during which ions move in a correlated manner remembering their original position, a
 stochastic regime when the initial vicinity of the ions is completely forgotten takes over. As
 we used the ions as tracers of magnetic field, we can talk about the stochasticity of magnetic field lines (as traced by ions).
 
 \section{D: Reconnection diffusion and collecting of matter along magnetic field lines}

Another concept that was discussed in the literature is the one based on changing the flux to mass ratio
by collecting matter along magnetic field lines. Similar to reconnection diffusion, this process
can happen within the MHD approximation and it does not depend on the degree of media ionization (see Vazquez-Semadeni et al. 2011). 
Incidentally, as one can easily observe in Figure 2, the motion along field lines is also
a part of the reconnection diffusion concept. The difference between the two aforementioned pictures arises, however, from the fact
that within the ``motion along magnetic field lines'' concept, flux freezing holds and the fluid moves along field lines
that preserve their identity through the process. On the contrary, the reconnection diffusion concept is based on fast reconnection
arising from the LV99 process or, equivalently, on the violation of flux freezing in turbulent flows that arises from 
fundamental properties of Lagrangian dynamics of these flows (see ELV11). 

It is beyond doubt that the collection of matter along magnetic field lines is an important process for laminar flows. Reconnection diffusion is not applicable for those as it requires turbulence for enabling both fast reconnection and diffusion
 of magnetic fields. However, as we mentioned earlier, observations testify that turbulence is ubiquitous in the ISM. 
 The fact that reconnection diffusion does act to smooth magnetic field lines twisted by turbulence does help promote motion of matter along field lines in turbulent media. However, it is necessary to investigate the relative importance of the concept of moving matter along {\it turbulent} magnetic field lines adopting the parameters of turbulence 
consistent with observations and appealing to the scalings arising from the widely accepted GS95 model 
of magnetic turbulence\footnote{The latter, while not being exact, is so far the
best model of MHD turbulence that we are aware of (see numerical testing in Beresnyak \& Lazarian 2010, Beresnyak 2011, 2012). The applicability of the GS95 scalings to the 
compressible turbulence was demonstrated via numerical simulations 
(Cho \& Lazarian 2002, 2003, Kowal \& Lazarian 2010).}. 

While the motion of fluid along laminar magnetic fields is well defined, the situation is a bit more complicated in turbulent fluids.
Thinking about the motion of fluids along turbulent magnetic fields one may visualize both the motion along the mean magnetic 
field and the motion along the actual turbulent field. Below we discuss both cases for the  
model cloud that we discussed in Lazarian et al. 2012, \S4. 

Consider first to what extent the actual fluid dynamics can be approximated by the motion
along the mean field. Assume that the material in the envelope is being collected along a
cylinder of $d_{env}=1$ pc oriented along the mean magnetic field. Taking the typical density of interstellar medium of 
$n_{ISM}=20$ cm$^{-3}$, one can estimate the length of the flux tube using the mass conservation relation 
\begin{equation}
d_{envelope} n_{envelope}\approx D_{tube, env} n_{ISM},
\end{equation}
which for the adopted value $n_{envelope}=10^3$ cm$^{-3}$ provides $D_{tube, env}=50$ pc. If interstellar densities are smaller, the collection distance $D_{tube, env}$ only increases. 
Assuming that the injection scale of interstellar turbulence is $L=100$ pc,
one has to accept that the magnetic field lines undergo Richardson superdiffusion that obeys Eq. (6) in Lazarian et al. (2012) (see also
ELV12 and ref. therein, Lazarian et al. (2004) for the numerical verification). Ignoring the initial separation
of the field lines $l_0\approx d_{env}$ one can estimate the separation as 
\begin{equation}
l^2_{diff}\approx (V_L^3/L) t^3,
\label{l2}
\end{equation}
where we took into account that the cascading rate $\alpha$ in MHD turbulence is $V_L^3/L$ if turbulence is injected
at the velocity $V_L=10$ km/s, which is larger than the Alfv\'{e}n velocity in the diffuse gas. 

In the scenario of gas being collected along magnetic field lines, the velocities relevant for creating 
molecular clouds are due to turbulence. The maximal turbulent velocity is limited to the injection turbulent velocity\footnote{In reality only a fraction of 
20\% of turbulent energy is in the potential motions that can be associated with compressions along magnetic field lines (see Cho \& Lazarian (2002)). In the text for obtaining the upper limit
for the compression, we consider the limiting case of the compressible velocity equal to the injection velocity.} that we denoted $V_L$. Therefore, the minimal time for the accumulation of the matter for the envelope is
\begin{equation}
t_{min}\approx D_{tube, env}/V_L.
\label{tmin}
\end{equation}
Combining Eqs. (\ref{l2}) and (\ref{tmin}) one gets
\begin{equation}
l^2_{diff}\approx D^3_{tube, env}/L
\label{l4}
\end{equation}
which for the adopted injection scale $L=100$ pc and estimated value of $D_{tube, env}$ provides the separation of
the field lines of $l_{diff}\approx 30$ pc. The fraction of material that actually moves along the the cylinder oriented along the mean magnetic field
in our case is $\sim d_{env}^2/l_{diff}^2\sim 10^{-4}$. 

In the case of the core one can assume that the mean magnetic field is constricted at the place of the core in accordance to magnetic flux
conservation:
\begin{equation}
B_{initial} d_{ini, core}^2=B_{final} d_{core}^2
\end{equation}
and we will use $d_{ini, core}$ as the diameter of the flux tube along which the collection of matter parallel to magnetic field is considered. Adopting
the value of $B_{initial}$ of 10 $\mu$G one gets $d_{ini, core}\approx 0.3$ pc. The same arguments that we used for the envelope provide for the core
the collection scale of 150 pc, which is larger than the assumed scale of the turbulence injection. As a result, spreading of the magnetic field
lines over this scale exceeds the injection scale $L$ and, compared to the envelope, an even smaller fraction of matter can be considered to move along
mean magnetic field.\footnote{Numerical simulations show some correlation between magnetic field and velocity in transAlv\'{e}nic runs (see Kritsuk et al. 2012). This by itself
cannot be used as an argument in favor of collecting matter in clouds along large insterstellar distances, as magnetic field lines are wandering and
spreading.} 

As the previous discussion shows that the collection of matter along the mean magnetic field may not be tenable for the parameters of turbulence that
we adopted, one may wonder whether it may be good to talk about collecting matter along the stochastic magnetic field lines but assuming that these
field lines are static and the diffusion of magnetic field during the collection time is negligible. Consider the envelope case, as it presents a less extreme
example of matter accumulation. It is easy to see that the issue of ``static'' is violated, as during the time of the collection, the magnetic field and fluid in the core performs $(n_{cloud}/n_{init}) (d_{envelope}/L)^{1/3}$ turnovers which is about 17 for the numbers adopted. The diffusion of magnetic field from the
core over the collection time obeys the Richardson law, i.e.
$\langle y^2 \rangle \approx V_L^3 t^3/L$,
which for the collection time $t_{min}$ provides the estimate that coincides with one given by Eq. (\ref{l2}). Therefore, the spread of magnetic field 
lines induced by the diffusion in turbulent media on the timescale of $t_{min}$ is very large compared to the scale of the envelope and we have to accept that magnetic field diffusion is very important. In other words, it is difficult to justify the collection of the matter along passive and static magnetic field lines. It is easy to see that the collection of matter within the core appealing to driving matter along turbulent magnetic field lines and ignoring the magnetic field line spread is even more questionable than for the case of the envelope. 

We should admit that our arguments above do not account for self-gravity, which may change the ability of magnetic field lines to spread at the core,
for instance. However, self-gravity cannot be important over the entire extent of flux tubes which is $\sim 50$ pc for the envelope and
$\sim 150$ pc for the core. Therefore we believe that, while matter is moving along magnetic field lines, the effects of reconnection diffusion cannot
be ignored. In fact, due to the pull of gravity we expect that matter in the presence of reconnection diffusion will be collected from all sides, rather 
than exclusively from the local direction of the magnetic field. Movement of matter along local field lines will of course be stronger than along other directions. In other words, the motion of matter is three dimensional within turbulent fluids and the notion of a one dimensional approximation assumed within the concept of the ``matter collection along magnetic field lines'' is of limited applicability if the present
day understanding of MHD turbulence is correct. This, in fact, corresponds well to the notion of violation of magnetic flux freezing in turbulent fluids (see
ELV11).

\begin{acknowledgments}
   AL acknowledges the support the NASA grant NNX09AH78G, NSF grant AST-1212096 the Vilas Associate Award as well as the NSF Center for Magnetic Self-Organization in Laboratory and Astrophysical Plasmas. The initial study by AL was performed with the support of the Humboldt Award at the Universities of Cologne and Bohum as well as benefited from his stimulating stay at the International Institute of Physics in Natal (Brazil). AL thanks Bruce Elmegreen for many stimulating discussions on the role of reconnection diffusion on star formation that substantially influenced our understanding of the processes.  In addition, 
AL thanks Greg Eyink, Alex Branderburg, Enrique Vazquez-Semadeni, Elisabeth de Gouveia dal Pino, Marcia Leao, Reinaldo Santos-Lima, Ethan Vishniac and Ellen Zweibel for useful exchanges. AE acknowledges support from CONACyT grants 101975 and 167611, as well as DGAPA-UNAM IN105312 grant. RC's 
  research is partially supported by grant NSF AST 10-07713. We thank the anonymous referee for helpful suggestions. 
\end{acknowledgments}

\end{document}